\documentclass[%
 aip,
 amsmath,amssymb,
reprint, twocolumns 
]{revtex4-1}

\usepackage{graphicx}
\usepackage{dcolumn}
\usepackage{bm}

\usepackage[utf8]{inputenc}
\usepackage[T1]{fontenc}
\usepackage{mathptmx}
\usepackage{etoolbox}
\usepackage{amsmath}
\usepackage{amsfonts}
\usepackage{amssymb}
\usepackage{epsfig}
\usepackage{epstopdf}
\usepackage{subfigure}
\usepackage{booktabs}
\usepackage{algorithmic}
\usepackage{color}
\usepackage{wasysym}
\usepackage{multirow}
\usepackage[ruled,vlined]{algorithm2e}
\newcommand{\RNum}[1]{\uppercase\expandafter{\romannumeral #1\relax}}

\newtheorem{Def}{Definition}

\makeatletter
\def\@email#1#2{%
 \endgroup
 \patchcmd{\titleblock@produce}
  {\frontmatter@RRAPformat}
  {\frontmatter@RRAPformat{\produce@RRAP{*#1\href{mailto:#2}{#2}}}\frontmatter@RRAPformat}
  {}{}
}%
\makeatother
\begin{document}

\preprint{AIP/123-QED}

\title[Sample title]{Synchronization of multiple rigid body systems: a survey}

\author{Xin Jin}
\email{jx\_9810@163.com}
	\altaffiliation{Xin Jin was with the Key Laboratory of Smart Manufacturing
	in Energy Chemical Process, Ministry of Education, East China
	University of Science and Technology, Shanghai 200237. Now he is with the Research Institute of
	Intelligent Complex Systems, Fudan University, Shanghai, China.}
\affiliation{The Research Institute of Intelligent Complex Systems, Fudan University, Shanghai 200433, China}%

\author{Daniel W. C. Ho}
 \email{madaniel@cityu.edu.hk}
\affiliation{The Department of Mathematics, City University of Hong Kong, Hong Kong, China
}%

\author{Yang Tang}
\homepage{Corresponding author: Yang Tang.}
 \email{yangtang@ecust.edu.cn}
\affiliation{The Key Laboratory of
 Smart Manufacturing in Energy Chemical Process, East China University of Science and Technology, Shanghai 200237, China}%

\date{\today}

\begin{abstract}
The multi-agent system has been a hot topic in the past few decades owing to its lower cost, higher robustness, and higher flexibility. As a particular multi-agent system, the multiple rigid body system received a growing interest for its wide applications in transportation, aerospace, and ocean exploration. Due to the non-Euclidean configuration space of attitudes and the inherent nonlinearity of the dynamics of rigid body systems, synchronization of multiple rigid body systems is quite challenging. This paper aims to present an overview of the recent progress in synchronization of multiple rigid body systems from the view of two fundamental problems. The first problem focuses on attitude synchronization, while the second one focuses on cooperative motion control in that rotation and translation dynamics are coupled. Finally, a summary and future directions are given in the conclusion.
\end{abstract}

\maketitle
\begin{quotation}
	The distributed sensing, decision-making, and cooperative control of multi-agent systems has been thoroughly investigated in the fields such as sensor networks, social networks, distributed computing, and robotics in the past decades. 
	Recently, multiple rigid body systems as a particular kind of multi-agent system attracted a lot of interest from researchers owing to its potential applications in aerospace engineering, unmanned vehicles, and industrial robotics. 
	This work aims to give a review of the recent research progress in synchronization of multiple rigid body systems from the aspects of two fundamental problems, which are attitude synchronization and coordination control of multiple rigid body systems. 
	The basic kinematic and dynamic model of describing rigid body systems are introduced, and the important results as well as the comparisons are given. Finally, several future topics are outlined. 
\end{quotation}
\section{Introduction}
\begin{figure*}[htb]	
	\centering
	\subfigure[]	
	{
		\includegraphics[scale=0.5]{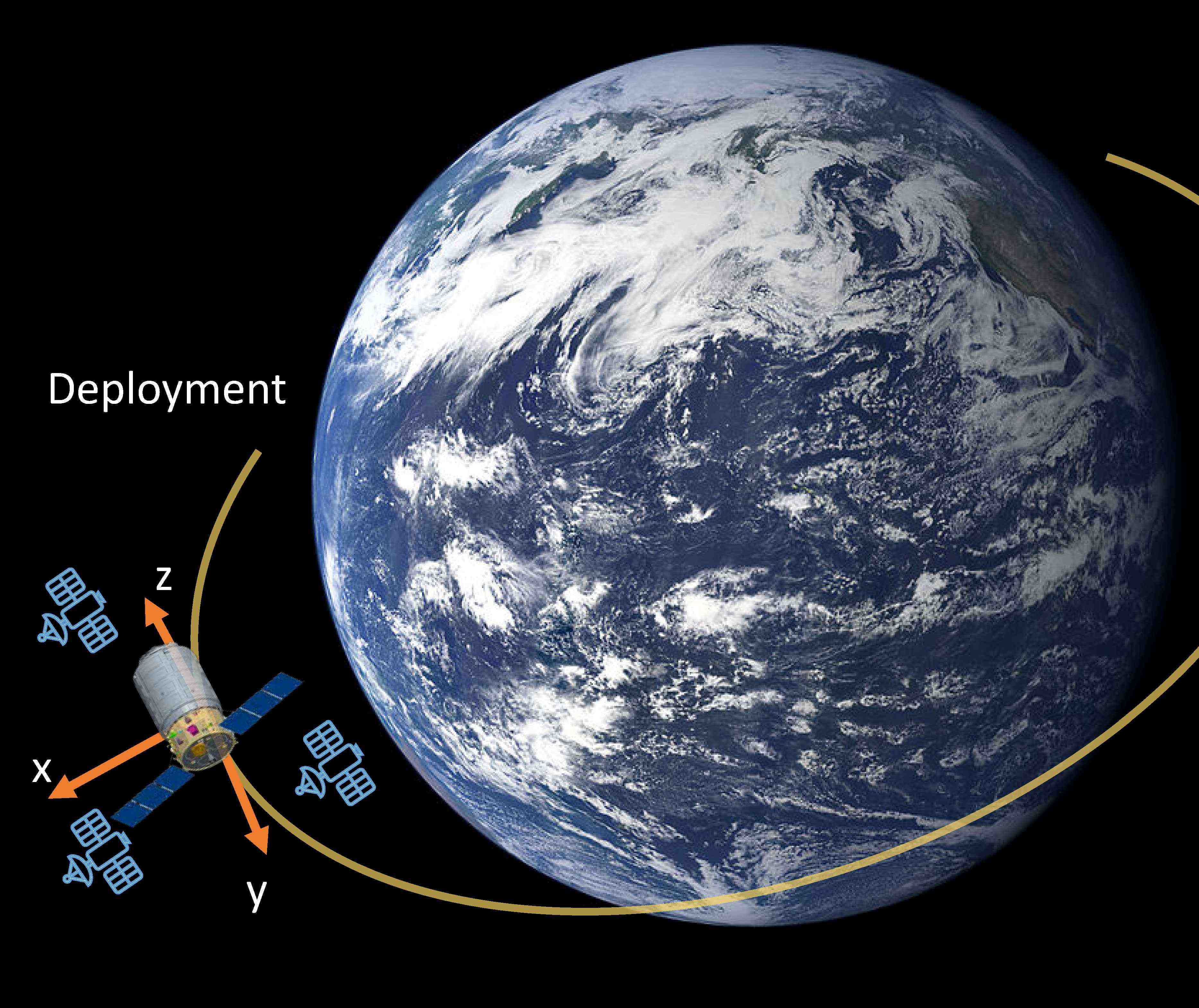}
	}	
	\subfigure[]
	{
		\includegraphics[scale=0.5]{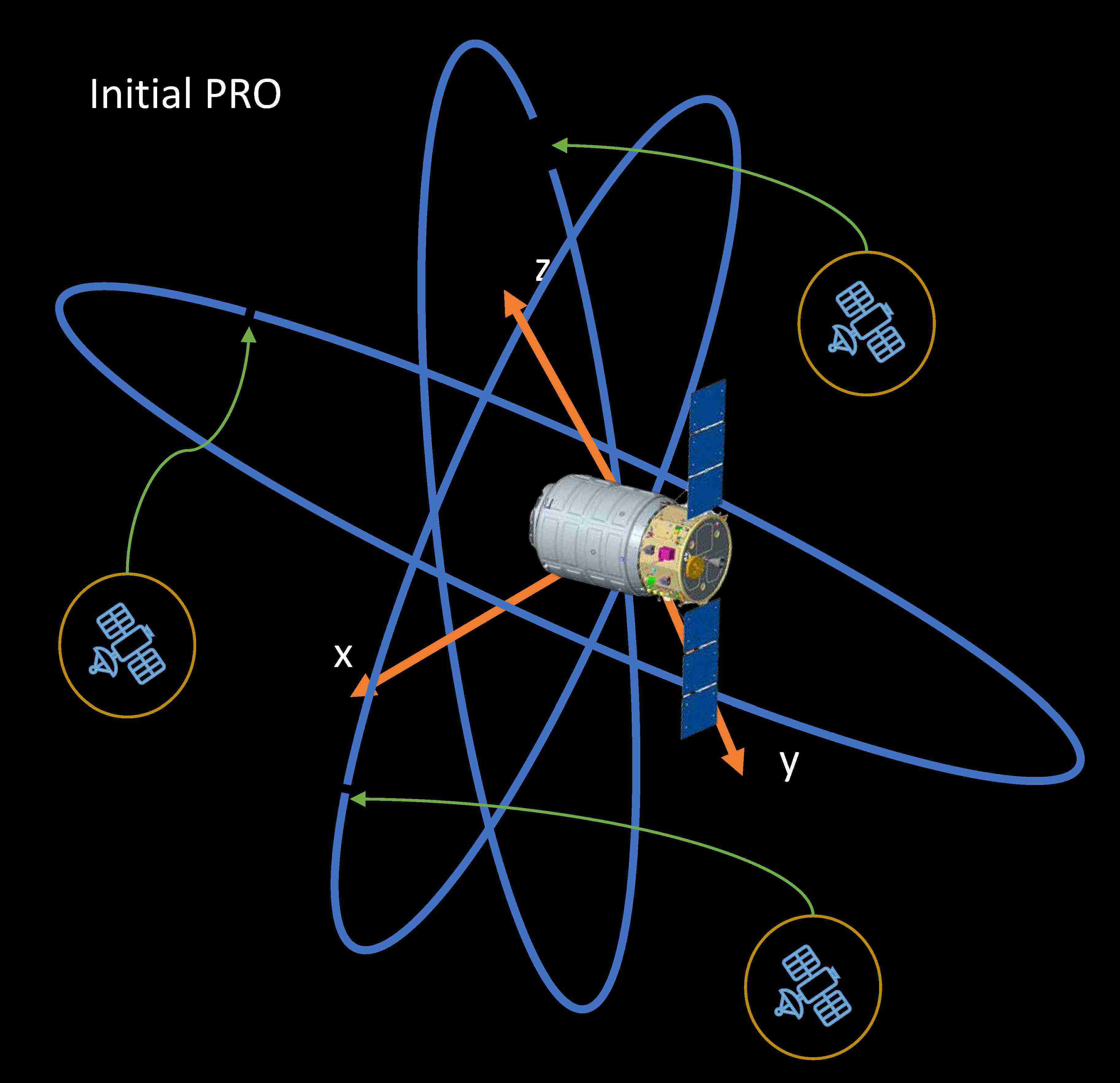}
	}	
	\subfigure[]
{
	\includegraphics[scale=0.5]{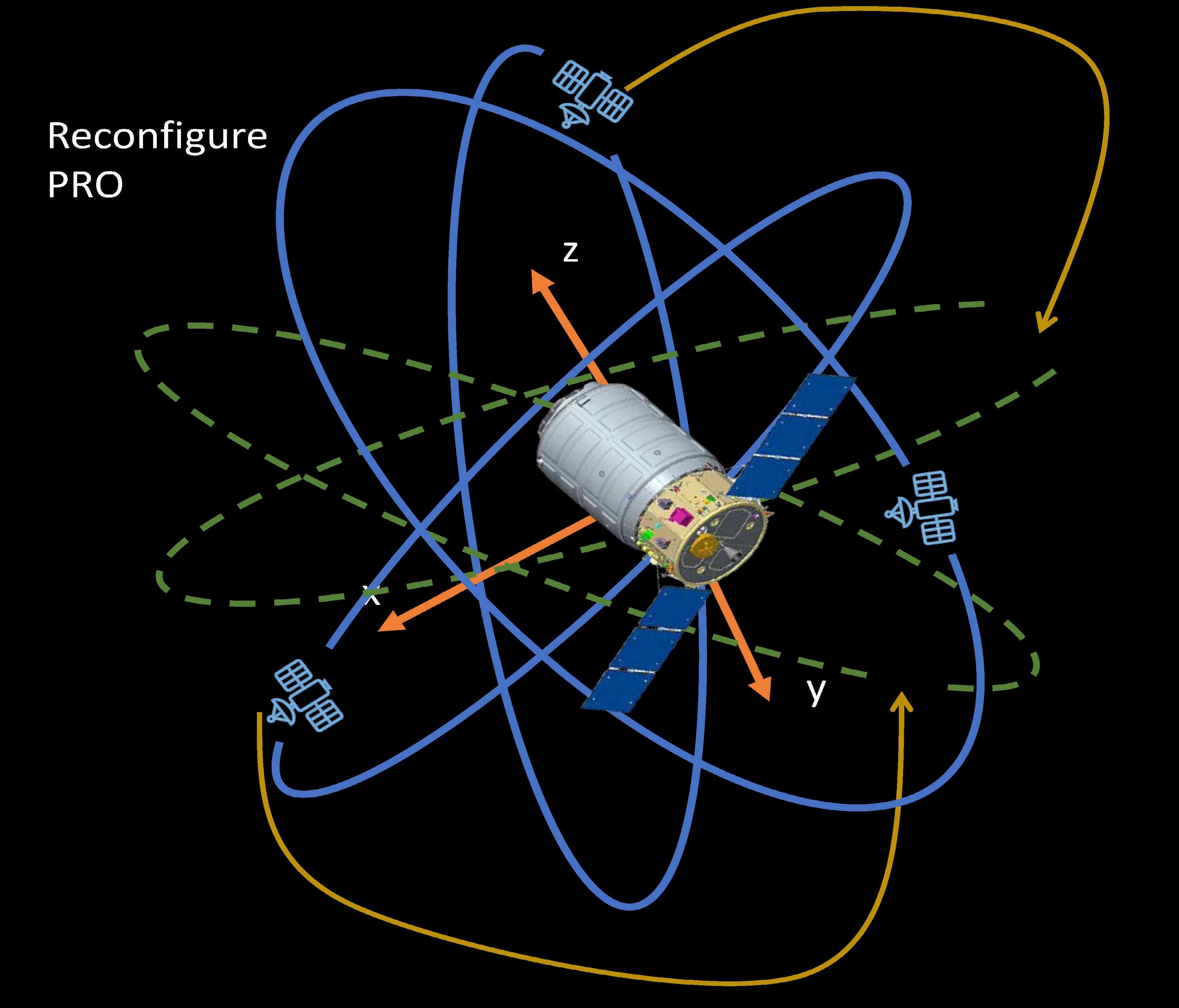}
}	
	\subfigure[]
{
	\includegraphics[scale=0.5]{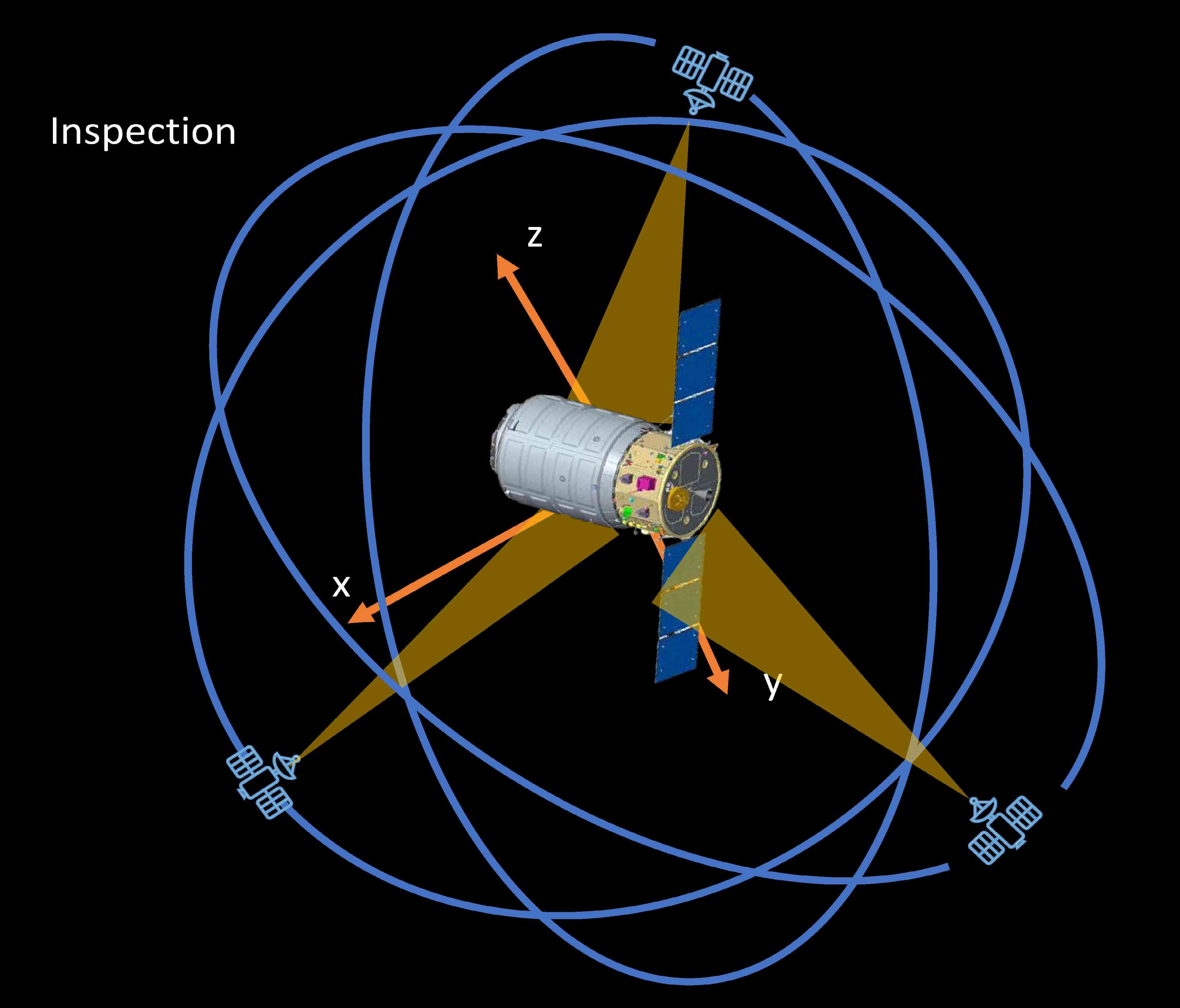}
}	
	\caption{An inspection mission with nano-satellites. (a). Deployment of three satellites around the target spacecraft. (b). Initialization of three satellites into stable relative orbits.
	(c). Reconfiguration of three satellites to observe a different area on the target.
(d). Coordinated pointing control for observing the target spacecraft \cite{2022_JGCD_Chung}. }		
	\label{observation}
\end{figure*}

{T}{}he last decades have witnessed a significant progress in consensus studies of multi-agent networks \cite{2005_ACC_W.Ren,2007_Proceeding_R.Olfati-Saber,2008_PR,2016_PR,2020_Chaos_Y.Tang,2020_TCYB_Y.Tang,2023_Chaos}. 
Through local knowledge and information interaction, agents can cooperatively complete a complicated task with lower cost, higher flexible, and higher robustness \cite{2008_Book_RenWei,2011_distributed_W.Ren}.   
As a kind of particular case in multi-agent systems, multiple rigid body systems have stronger engineering backgrounds in the fields such as robotics \cite{2009_TRO_JJE}, satellites \cite{2022_JGCD_Chung}, and unmanned aerial vehicles \cite{2012_TAES_uav}.
For example, with the rapid development of the perception ability of autonomous systems and artificial intelligence, in aerospace engineering, the coordination of a group of small or nano-satellites can deliver a comparable or even stronger capability compared with a monolithic satellite in completing the complex space mission such as distributed observation, on-orbit assembly, and asteroid defense \cite{2020_Patterns,2022_TNNLS_survey}. 
Typically, an inspection mission for coordinated observing a spacecraft target with a group of nano-satellites is shown in Fig. \ref{observation} \cite{2022_JGCD_Chung}.

The rigid body can be considered as an idealization model of a body that does not deform or change shape under external forces \cite{2001_mechanics}. 
The motion of rigid body is composed of the rotational and translational motion.
From this point of view, synchronization of multiple rigid body systems in the literature can be divided into two categories. The one is attitude consensus which focuses on the rotational motion, while the other focuses on the coordinated motion control in which the rotational and translational motion are coupled together. 
Note that the "synchronization" and the similar noun "consensus" are usually regarded as exchangeable concepts in the literature,  which have a common intension. 
Thus, we use these two words with no distinction in this paper.

{\color{black}For the rotational motion, the attitude representation of a rigid body can be classified into parameterized representations and rotation matrices.
The parameterized representations include Euler angles \cite{2019_TAES_J.Shan}, Rodrigues parameters \cite{2020_Auto_X.Jin}, Modified
Rodrigues parameters (MRPs) \cite{2010_Auto_Z.Meng_attitude}, and unit quaternions \cite{2011_ControlSystemMagazine}.
Euler angles are widely used in the formation control of aerial vehicles and quadrotors based on linearized models \cite{2013_TCYB_Tangyang_sychronization}. 
Note that the underlying space of rotation matrices $\mathbb{SO}(3)$ is non-diffeomorphic to any Euclidean spaces \cite{1993_JAS_M.Shuster_attitude}. 
Euler angles and modified
Rodrigues parameters (MRPs) evolving on Euclidean spaces only achieve local convergence due to the singularity problem \cite{2010_TCST_W.Ren_AttitudeSynchronization}. 
Several works studied the attitude synchronization problem based on the unit quaternion since it can globally represent the attitude. 
Nevertheless, it may suffer from the unwinding phenomenon due to the non-uniqueness of describing attitudes \cite{1993_JAS_M.Shuster_attitude}. 
Hence, to overcome this problem, a hybrid feedback control approach has been utilized to design the controller, which can achieve the global attitude synchronization \cite{2012_TAC_A.R.Teel}. }

The rotation matrix is the only global and unique attitude representation, which completely represents the attitude. 
Motivated by this fact, a substantial amount of literature has been devoted to the investigation of attitude consensus based on rotation matrices \cite{2009_Auto_Sarlette,2009_TAC_Spong,2019_TCYB_Z.Meng}.
However, due to the geometric topological constraints, there is no continuous time-invariant feedback which can achieve the global attitude synchronization \cite{2000_SCL_Bhat}.
Therefore, many studies have focused on considering the almost global attitude synchronization  \cite{2014_Auto_J.T_attitude,2018_TAC_JM_n-sphere} and see the references therein.
It is noteworthy that, in some scenarios, the incompleted or reduced attitude consensus has a more relevant practical application such as coordinated pointing of nano-satellite \cite{2017_Auto_Dimos}. 
The completed attitude consensus and incompleted consensus can be both considered in a unified framework on $\mathbb{S}^{n},\; n=2,3$\cite{2020_TAC_Dimos_attitude}.
It should be noted that the synchronization problem on $\mathbb{S}^{n}$ attracts many interests from different disciplines, including coupled oscillators \cite{2004_PRE,2011_PRE}, complex networks \cite{2021_Chaos,2022_Chaos,2023_PR}, and quantum mechanics \cite{2015_tac_Mazzarella,2016_TAC_Shi,2017_tac_Shi}.
More generally, the underlying space $\mathbb{S}^{2}$, $\mathbb{S}^{3}$, and $\mathbb{SO}(3)$ can be considered as a Riemannian manifold \cite{2011_ARC}. 
Due to the geometric topology constraint, the consensus protocol is quite difficult to design and the convergence domain is hard to be determined analytically on nonlinear spaces.
Recently, there are some novel approaches dealing with the consensus problem on different Riemannian manifolds such as gradient flow methods \cite{2021_TAC_JM,2020_Auto_J.Thunberg}, lifting methods \cite{2018_Auto_Thunberg}, and matrix decomposition methods \cite{2018_Auto_JT_Dynamic}.

{\color{black}Rigid body systems' rotational and translational motions are usually coupled in dynamic models.
A rigid body dynamic model is challenging to be obtained precisely in practical applications, especially when it contains unknown dynamics and environmental disturbances \cite{2015_TCNS_Dixon}.   
The Euler-Lagrange equation is equivalent to Newton's laws of motion in classical mechanics.
It is an effective method to describe the rigid body dynamics when the force vectors are particularly complicated \cite{2001_mechanics}. 
In the past few decades, the coordination control of Euler-Lagrange systems has been widely studied in the literature under undirected graphs \cite{2009_IJC_W.Ren_EL} and directed graphs \cite{2013_Auto_H.Wang_EL,2012_TRO_Chopra_EL,2014_TAC_H.Wang_EL}, respectively.}
{\color{black}Lately, motivated by the fact that communication among agents is unreliable in real applications, the coordination control of networked Euler-Lagrange systems is considered with time-delays \cite{2014_TAC_Abd}, packet dropouts \cite{2017_TAC_Abdessameud_EL}, and sampled-data mechanism \cite{2018_TII_Zhangwb_Sampled-dataEL}.
In addition, as a particular case of non-periodic sampled-data setting, the event-triggered coordination control has also been extensively studied for Euler-Lagrange systems to reduce the communication cost \cite{2019_TNSE_X.Jin,2022_TCST_X.Jin}.

The most of results on the coordination control of Euler-Lagrange systems are based on the fundamental properties of Euler-Lagrange dynamics such as anti-symmetry and parameterized linearity, which are quite ideal. For example, when considering the motion of a rigid body on a special Euclidean group $\mathbb{SE}(3)$, the anti-symmetric property of Euler-Lagrange dynamics may not be guaranteed \cite{2019_Auto_Dimos}. 
Extensive research has been conducted on the coordinated motion control on $\mathbb{SE}(3)$ due to its
theoretical challenges in handling the nonlinear configuration space \cite{2012_TAC_Spong} and the switching topologies \cite{2016_Auto_J.Thunberg}.}

Overall, synchronization of multiple rigid body systems have been thoroughly investigated in the past decades and have attracted a growing interest for researchers in theoretical research as well as in practical applications.
Up to our knowledge, very few works give a comprehensive literature review for synchronization of multiple rigid body systems.  
Compared with the recent survey on attitude consensus of multiple spacecraft \cite{2022_Chen}, we further consider a more general multiple rigid body model where the rotational dynamics and translational dynamics may couple together.

The remaining part of this paper proceeds as follows. Section \RNum{2} presents the notation and preliminary knowledge, including graph theory, attitude representations, rigid body kinematics, and rigid body dynamics. Sections \RNum{3} and \RNum{4} show the representative results of attitude synchronization and coordination control of multiple rigid body systems. 
The conclusion is drawn in Section \RNum{5} finally. 
\section{Preliminaries }
\subsection{Notations}
$\mathbb{R}^{N}$ and $\mathbb{R}^{{N}\times {N}}$ represent the Euclidean vector space and real matrix space, where $N$ is a positive integer number. $\mathbb{N}^{+}$ denotes the positive integer numbers.  $\mathbb{R}^{+}$ represents the positive real numbers.
For a vector $\mathbf{x}=[x_1,...,x_N]^{\top}\in \mathbb{R}^{N}$, $\|\mathbf{{x}}\|$ denotes the Euclidean norm, which is defined as $\|\mathbf{x}\|=\sqrt{x_{1}^{2}+x_{2}^{2}+...+x_{N}^{2}}$. 
{$\mathbf{0}_{3}$ denotes a zero vector in $\mathbb{R}^{3}$.}
$\bar{\lambda}(\mathbf{A})$ and $\underline{\lambda}(\mathbf{A})$ denote the maximum and minimum eigenvalue of the matrix $\mathbf{A}$, respectively.
$\text{tr}(\mathbf{A})$ represents the trace of matrix $\mathbf{A}$.
$|\mathcal{D}|$ denotes the number of elements in $\mathcal{D}$. 
The set $\mathbb{SO}(3)$ is a special orthogonal group and the set $\mathfrak{so}(3)=\{\mathbf{X}\in \mathbb{R}^{3 \times 3}:\mathbf{X}^{\top}=-\mathbf{X}\}$.
The operators $(\cdot)^{\wedge}$ and $(\cdot)^{\vee}$ represent a mapping between the vector {\color{black}$\mathbf{x}=[x_{1},x_{2},x_{3}]^{\top}$ and the skew symmetric matrix $\mathbf{X}=
	\left[\begin{matrix}
		0&-x_{3}& x_{2}\\
		x_{3}&0& -x_{1}\\
		-x_{2}&x_{1}&0\\
	\end{matrix}\right]
	$, where $\mathbf{x}^{\wedge}=\mathbf{X}$ and $\mathbf{x}=\mathbf{X}^{\vee}$. }
{\color{black}	The sign function is defined as  $ \operatorname{sgn}(x)=
	\left\{\begin{matrix}
		1&x>0\\
		0&x=0\\
		-1&x<0\\
	\end{matrix}	\right.
	$.}
\subsection{Graph Theory}
We first introduce some basic concepts in graph theory.
Let $\mathcal{G}=\mathcal{G}\{\mathcal{V},\mathcal{E}\}$ denote a topology graph, in which $\mathcal{V}=\{1,...,N\}$ represents  a node set, and $\mathcal{E}\subseteq \mathcal{V}\times \mathcal{V}$ represents an edge set. 
{The edge denoted as $(i,j)\in \mathcal{E}$ means that the node $i$ is the node $j$'s neighbor.} In other words, node $j$ can receive node $i$'s message. 
All the neighbors of node $i$ forms a set denoted as   $\mathcal{N}_i=\{j\in\mathcal{V}: (j,i)\in\mathcal{E}\}$. 
For one undirected graph, if $(i,j)\in \mathcal{E}$, then $(j,i)\in \mathcal{E}$. 
A graph is connected if there exists a link between any two nodes.
Let an adjacency matrix $\mathcal{A}=[a_{ij}]\in \mathbb{R}^{N\times N}$ associate with the graph $\mathcal{G}$,
where $a_{ij}>0$ if the node $j$ is the node $i$'s neighbor and zero otherwise. 
{We suppose $a_{ii}=0$, which means that the self-connection is excluded here.}
The Laplacian matrix is denoted by $\mathcal{L}=[l_{ij}]\in \mathbb{R}^{N\times N}$. The diagonal elements $l_{ii}$ of the Laplacian matrix are the in-degree of node $i$, which can be calculated as $\sum_{j=1}^{N}a_{ij}$, and the non-diagonal elements are defined as $l_{ij}=-a_{ij}, i \neq j$. 
 
 {\color{black}Since the topology graph can be time-varying in practical multi-agent systems, we introduce some definitions of switching topologies. 
Denote all the possible topologies of the graph $\mathcal{G}$ as $\mathcal{G}_{1},\mathcal{G}_{2},...,\mathcal{G}_{M}$. 
Define a continuous piecewise constant switching signal as $\sigma(t):[0, \infty) \rightarrow\{1,2, \ldots, N\}$.
Let a dwell time $\tau_D>0$ be as a lower bound between any two consecutive switching times, which are the switching instances $\left\{\tau_k \mid k=1,2, \ldots\right\}$ satisfying
\begin{align}
	\inf _k\left(\tau_{k+1}-\tau_k\right) \geq \tau_D.
\end{align}
The union graph of $\mathcal{G}_{\sigma(t)}$ during the time interval $\left[t_1, t_2\right)$ is defined as 
 $$\mathcal{G}\left(\left[t_1, t_2\right)\right)=\bigcup_{t \in\left[t_1, t_2\right)} \mathcal{G}_{\sigma(t)}=\left(\mathcal{V}, \bigcup_{t \in\left[t_1, t_2\right)} \mathcal{E}_{\sigma(t)}\right),$$
where $t_1<t_2 \leq+\infty$.
A wild assumption of switching topology is called jointly connected, which is defined as follows.
\begin{Def}\label{switching}
	The switching topology  $\mathcal{G}_{\sigma(t)}$ is jointly connected if there exists a constant $T>0$ such that the union graph $\mathcal{G}\left[t, t+T\right)$ is connected for any $t\geq0$.
\end{Def}}

\subsection{Attitude representations and kinematics}
Let $\mathcal{F}_{w}$ denote the world frame and $\mathcal{F}_{i}$ the local body frame of the rigid body $i$, where $i={1,...,N}$.
The attitude of each rigid body $i$ in the world frame $\mathcal{F}_{w}$ is denoted by $\mathcal{R}_{i}\in \mathbb{SO}(3)$. 
Some examples of attitude representations and their kinematics are summarized in the following content.
The global and unique property of attitude representations are summarized in Table \ref{attitude_representations}.

\subsubsection{Rotation matrices}
\begin{figure}[h]	
	\centering
	{
		\includegraphics[scale=0.6]{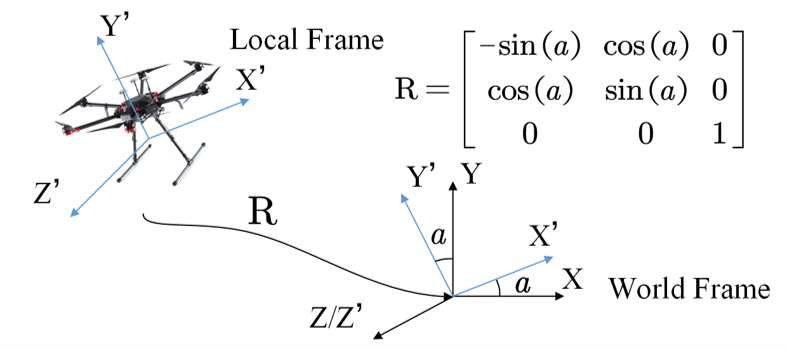}
		\caption{Rotation between the local and world frame.}	\label{rotation}
	}
\end{figure}
The rotation matrix is a linear transformation, which describes the rotation between the local frame and the world frame as shown in Fig. \ref{rotation}.
All the rotation matrices form a special rotation group as follows,
\begin{align}
	\mathbb{SO}(3)=\left\{\mathcal{R}_{i} \in \mathbb{R}_{i}^{3 \times 3} \mid \operatorname{det} \mathcal{R}_{i}=1, \mathcal{R}_{i} \mathcal{R}_{i}^{\top}=\mathcal{R}_{i}^{\top} \mathcal{R}_{i}=\mathbf{I}_{3}\right\}.
\end{align}
Based on the rotation matrix, the attitude kinematics of the $i$th rigid body is governed by 
\begin{align}
	\dot{\mathcal{R}}_i=\mathcal{R}_i {\boldsymbol{\omega}}_i^{\wedge},\; i\in\mathcal{N},
\end{align}
where $\mathcal{R}_{i}\in \mathbb{SO}(3)$, and $\boldsymbol{\omega}_i\in \mathbb{R}^{3}$ is the angular velocity. 
\begin{table}[htb]
	\caption{Properties of attitude representations \cite{2011_ControlSystemMagazine}.}
	\label{attitude_representations}
	\begin{tabular}{l|c|c}
		\hline
		Attitude representations      & Global & Unique \\ \hline
		Euler angles                  & No     & No     \\ \hline
		Rodrigues parameters          & No     & No     \\ \hline
		Modified Rodrigues parameters \quad \quad\quad\quad\quad& No     & No     \\ \hline
		Quaternions                   & Yes    & No     \\ \hline
		Rotation matrices             & Yes    & Yes    \\ \hline
	\end{tabular}
\end{table}
\subsubsection{Axis-angle representations}
Let $\mathbf{x}_{i}\in \mathbb{R}^{3}$ denote the axis-angle attitude representation of each rigid body $i$, which can be obtained by 
\begin{align}
	\mathbf{x}_{i}^{\wedge}=&\log\mathcal{R}_{i}\nonumber\\
	=&\theta_{i}{\mathbf{u}_{i}^{\wedge}},
\end{align}
where $\log: \mathbb{SO}(3)\rightarrow\mathfrak{so}(3) $ is the logarithm map, $\theta_{i}\in [0,\pi]$ is the rotation angle with respect to the rotation axis $\mathbf{u}_{i}\in \mathbb{R}^{3}$.

More specifically,
\begin{align}\label{u}
	\mathbf{u}^{\wedge}_{i}=\frac{1}{2\sin(\theta_{i})}(\mathcal{R}_{i}-\mathcal{R}_{i}^{\top})
\end{align}  and
\begin{align}\label{theta0}
	\cos(\theta_{i})=\frac{\text{tr}(\mathcal{R}_{i})-1}{2}.
\end{align} 
Note that the axis-angle vector is a global attitude representation, however, a pair of axis-angles corresponds to the same attitude at the point when $\theta_{i}=\pi$.

Based on the axis-angle representation, the attitude kinematics is given by \cite{2014_Auto_J.T_attitude} in the following,
\begin{align}\label{dx}
	\dot{\mathbf{x}}_{i}=\mathbf{J}_{\mathbf{{x}}_i}\bm{\omega}_{i}.
\end{align}
 The Jacobian matrix  $\mathbf{J}_{\mathbf{{x}}_i}$ is defined as
\begin{align}{\label{J}}
	\mathbf{J}_{\mathbf{{x}}_i}=&\mathbf{I}_{3}+\frac{\theta_{i}\mathbf{u}_{i}^{\wedge}}{2}+\Big(1-\frac{\theta_{i}}{2}\cot\frac{\theta_{i}}{2}\Big)(\mathbf{u}_{i}^{\wedge})^{2}\nonumber\\
	=&{{\mathbf{J}}^{\wedge}_{\mathbf{x}_{i}}}+\frac{\theta_{i}\mathbf{u}_{i}^{\wedge}}{2},
\end{align}  
where ${{\mathbf{J}}^{\wedge}_{\mathbf{{x}}_i}}=\mathbf{I}_{3}+\Big(1-\frac{\theta_{i}}{2}\cot\frac{\theta_{i}}{2}\Big)(\mathbf{u}_{i}^{\wedge})^{2}$ is the symmetric matrix.
{
	We know that if $\theta_{i}\in [0,\pi]$, the Jacobian matrix is positively definite, i.e., $\mathbf{x}_i^{\top}\mathbf{J}_{\mathbf{x}_i}\mathbf{x}_i>0$ for all $\mathbf{x}_i\neq \mathbf{0}_{3},  \mathbf{x}_i\in \mathbb{R}^3$. When $\theta_{i}=0$, ${{\mathbf{J}}^{\wedge}_{\mathbf{{x}}_i}}=\mathbf{I}_{3}$.}
In addition, we can get the geometric property of the Jacobian matrix in which the second and the third term 
are perpendicular to $\mathbf{x}_{i}$, i.e., $\mathbf{x}_{i}^{\top}\mathbf{J}_{\mathbf{{x}}_i}=\mathbf{x}_{i}^{\top}$. 
\subsubsection{Rodrigues parameters}
Let $\mathbf{y}_{i}\in \mathbb{R}^{3}$ denote the Rodrigues parameter attitude representation of each rigid body $i$, which is given by
\begin{align}\label{y}
	\mathbf{y}_{i}=\tan \frac{\theta_{i}}{2}\mathbf{u}_{i},
\end{align}
where $\mathbf{u}_{i}$ and $\theta_{i}$ are calculated by (\ref{u}) and (\ref{theta0}).
Note that the Rodrigues parameters have singularities when the rotation angle $\theta_{i}=\pm \pi$.
The attitude kinematics based on the Rodrigues parameters is given by 
\begin{align}
	\dot{\mathbf{y}}_i=\frac{1}{2} \mathbf{h}_i\left(\mathbf{y}_i\right) \bm{\omega}_i,
\end{align} 
where $\mathbf{h}_i\left(\mathbf{y}_i\right)=\mathbf{I}_3+\mathbf{y}_i^{\wedge}+\mathbf{y}_i \mathbf{y}_i^{\top}$.

\subsubsection{Modified Rodrigues parameters}
Let $\bm{\sigma}_{i}\in \mathbb{R}^{3}$ denote the modified Rodrigues parameter of $i$th rigid body, which is given by 
\begin{align}\label{MRP}
	\bm{\sigma}_{i}=\tan \frac{\theta_{i}}{4}\mathbf{u}_{i},
\end{align}
where $\mathbf{u}_{i}$ and $\theta_{i}$ are consistent with (\ref{y}). 
{\color{black}The Modified Rodrigues parameters also have singularity problems when $\theta_{i}=\pm 2\pi$. 
The attitude kinematics based on the Modified Rodrigues parameters is given by 
\begin{align}
	\dot{\boldsymbol{\sigma}}_i=\boldsymbol{G}\left(\boldsymbol{\sigma}_i\right) \boldsymbol{\omega}_i, \; i\in \mathcal{N},
\end{align}
where $\boldsymbol{G}\left(\boldsymbol{\sigma}_i\right)=\frac{1}{2}\left(\frac{1-\boldsymbol{\sigma}_i{ }^{\top} \boldsymbol{\sigma}_i}{2} \mathbf{I}_3+{\boldsymbol{\sigma}}_i^{\wedge}+\boldsymbol{\sigma}_i \boldsymbol{\sigma}_i^{\top}\right) \in \mathbb{R}^{3 \times 3}$. The matrix $\boldsymbol{G}\left(\boldsymbol{\sigma}_i\right)$ satisfies $\boldsymbol{G}\left(\boldsymbol{\sigma}_i\right) \boldsymbol{G}\left(\boldsymbol{\sigma}_i\right)^{\top}=\left(\frac{1+\boldsymbol{\sigma}_i^{\top} \boldsymbol{\sigma}_i}{4}\right)^2 \mathbf{I}_3$ \cite{1993_JAS_M.Shuster_attitude}.
\subsubsection{Unit quaternions}
Let $\bm{q}_{i}$ denote the unit quaternion of $i$th rigid body, which is defined as 
\begin{align}
	\bm{q}_{i}=\left[\begin{array}{l}
		\eta_{i} \\
		\bm{\epsilon}_{i}
	\end{array}\right] \in \mathbb{S}^3,
\end{align}
where $\mathbb{S}^3=\left\{(\eta_{i}, \bm{\epsilon}_{i}) \in \mathbb{R} \times \mathbb{R}^3: \eta_{i}^2+\bm{\epsilon}_{i}^{\top}\bm{\epsilon}_{i}=1\right\}$, $\eta_{i}\in \mathbb{R}$ is a scalar part, and $\bm{\epsilon}_{i}\in \mathbb{R}^{3}$ is a vector part. Each unit quaternion $\bm{q}_{i}\in 
\mathbb{S}^{3}$ has the inverse $\bm{q}_{i}^{-1}=(\eta_{i},-\bm{\epsilon}_{i})$. 
Note that a pair of antipodal unit quaternions $\pm\bm{q}_{i}\in \mathbb{S}^{3}$ corresponds to the same attitude $\mathcal{R}_{i}\in \mathbb{SO}(3)$. 
The quaternion kinematic equation for agent $i$ satisfies
\begin{align}\label{quaternion_kinematic}
	\dot{\bm{q}}_i=\frac{1}{2}\left[\begin{array}{c}
		-\bm{\epsilon}_i^{\top} \\
		\eta_i \mathbf{I}_{3}+{\bm{\epsilon}}_i^{\wedge}
	\end{array}\right] \bm{\omega}_i.
\end{align}}

\subsection{Parameterized attitude representations}
	The attitude representations in $\mathbb{R}^{3}$ are also called parameterized attitude representations. 
	In fact, the parameterized attitude representations can be considered as coordinates in a chart, which covers an open ball around the identity matrix on $\mathbb{SO}(3)$ \cite{2016_Auto_J.Thunberg}. 
	To make this point clear, a diffeomorphism mapping is used to give a unified definition for the parameterized attitudes \cite{2016_Auto_J.Huang_Attitude}. 
	
{\color{black}	Let 	
		$f: \mathcal{B}_{r}(\mathbf{I}) \subset \mathbb{SO}(3) \rightarrow \mathcal{B}_{r^{\prime}, 3}(0) \subset \mathbb{R}^{3}
	$ be defined as a diffeomorphism mapping from $\mathbb{SO}(3)$ to $\mathbb{R}^{3}$, where $\mathcal{B}_{r}(\mathbf{I})=\{\mathcal{R}\in \mathbb{SO}(3): d(\mathcal{R},\mathbf{I})<r\}$ is an open geodesic ball around the identity matrix in $\mathbb{SO}(3)$ with radius $r\leq \pi$, $\mathcal{B}_{r'}(\mathbf{0})=\{\mathbf{x}\in \mathbb{R}^3: \|\mathbf{x}\|<r'\}$ is an open ball around point $\mathbf{0}$ in $\mathbb{R}^{3}$. {If $r=\pi$, then $\mathcal{B}_{r}(\mathbf{I})$ covers $\mathbb{SO}(3)$ almost globally}, i.e., the set $\mathbb{SO}(3)-\mathcal{B}_{\pi}(\mathbf{I})$ has measure zero. In addition, if $r \leq \frac{\pi}{
		2}$, $\mathcal{B}_{r}(\mathbf{I})$ is convex. 
	Then, a general form of $f$ can be given as $   f(\mathcal{R})=g(\theta(\mathcal{R}))\mathbf{u}(\mathcal{R})$, where $\theta(\mathcal{R})$ is the geodesic distance between $\mathbf{I}$ and $\mathcal{R}$ denoted as $d(\mathbf{I}, \mathcal{R})=\theta$, $\mathbf{u}(\mathcal{R})\in \mathbb{S}^{2}$ is a unit vector which represents the rotation axis of $\mathcal{R}$, $g(\cdot): (-r, r) \rightarrow \mathbb{R}$ is an odd and strictly increasing function such that $f$ is a diffeomorphism.  
	Note that $r\leq \pi$ is the largest radius such that $f$ is a diffeomorphism. $\theta(\mathcal{R})$, $\mathbf{u}(\mathcal{R})$ can be obtained using the logarithm map in (\ref{u}) and (\ref{theta0}).

	Figure \ref{ball} illustrates the geometric configuration spaces of $\mathbb{SO}(3)$ and the open ball $\mathcal{B}_{\pi}(\mathbf{0}_{3})$. 
	Figure \ref{ball1} shows a diffeomorphism of $\mathbb{SO}(3)$, which is a solid closed ball $\bar{\mathcal{B}}_{\pi}(\mathbf{I}_{3})=\{\mathcal{R}\in \mathbb{SO}(3): d(\mathcal{R},\mathbf{I})\leq \pi \}$ with antipodal surface points identified. 
	Figure \ref{ball2} illustrates that the open ball  ${\mathcal{B}}_{\pi}(\mathbf{I}_{3})$ is diffeomorphic to the open ball ${\mathcal{B}}_{\pi}(\mathbf{0}_{3})$ which is embeded in Euclidean spaces.
\begin{figure}[htb]	
	\centering
	\subfigure[]	
	{
		\includegraphics[scale=0.38]{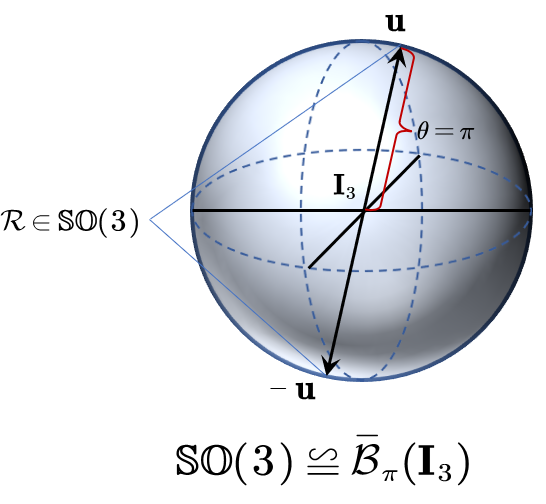}
		\label{ball1}
	}	
	\subfigure[]
	{
		\includegraphics[scale=0.38]{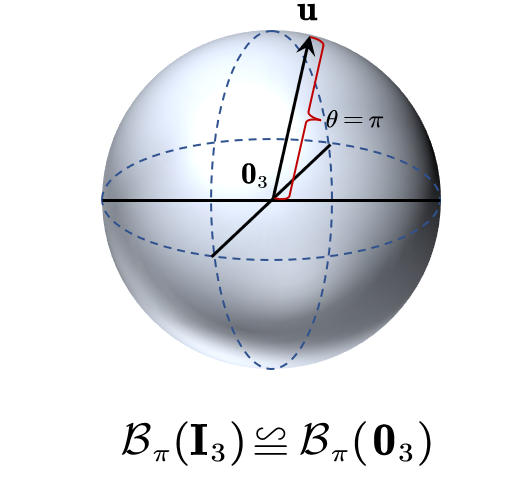}
		\label{ball2}
	}	
	\caption{The geometric configuration of $\mathbb{SO}(3)$ and  $\mathcal{B}_{\pi}(\mathbf{0}_{3})$. (a). A solid closed ball $\bar{\mathcal{B}}_{\pi}(\mathbf{I}_{3})$ with antipodal surface points identified. (b). An open ball ${\mathcal{B}}_{\pi}(\mathbf{I}_{3})$ diffeomorphic to the open ball ${\mathcal{B}}_{\pi}(\mathbf{0}_{3})$.}		
		\label{ball}
\end{figure}
}
		\begin{table}[]
				\caption{Local attitude representations\cite{2016_Auto_J.Thunberg}}
				\label{mapping}
		\centering
		{%
		\begin{tabular}{l|l|l|l}
			\hline
			Coordinates      & $f=g(\theta)\mathbf{u}\quad\quad$            & $r \quad\quad$             & $r' \quad\quad$   \\ \hline
			Axis-angles                             & $g(\theta)=\theta$                 & $\pi$           & $\pi$    \\ \hline
			{Rodrigues parameters} & $g(\theta)=\tan(\frac{\theta}{2})$ & $\pi$ & {$+\infty$}           \\ \hline
			Modified Rodrigues parameters \quad \quad& $g(\theta)=\tan(\frac{\theta}{4})$ & $\pi$ & 1                   \\ \hline
			$(\mathcal{R}-\mathcal{R}^{\top})^{\vee}$ & $g(\theta)=\sin(\theta)$           & $\frac{\pi}{2}$ & 1      \\ \hline
			Unit quaternions (Vector part)          & $g(\theta)=\sin(\frac{\theta}{2})$ & $\pi$           & 1      \\ \hline
		\end{tabular}%
	}
	\end{table}
     Based on the general form of $f$, the parameterized attitude representations and their corresponding configuration spaces in $\mathbb{R}^{3}$ are given in Table \ref{mapping}.
 
\subsection{Rigid body dynamics}
There are generally two approaches that describe the  dynamics of rigid body systems: Newton-Euler equations and Euler-Lagrange equations. 
\subsubsection{Newton-Euler equation}
The Newton-Euler equation is established by Newton's law of motion combing with the Euler equation for the rotational motion as follows,
\begin{align}\label{dynamics_newton}
	m_{i}\dot{\mathbf{v}}_{i}=&\mathbf{f}_{i},\nonumber\\
	\mathbf{J}_i \dot{\bm{\omega}}_i=& -{\bm{\omega}}_i^{\wedge}\mathbf{J}_i \bm{\omega}_i+\bm{\tau}_i,
\end{align}
where $m_{i}\in\mathbb{R}$ is the mass, $\mathbf{J}_{i}\in \mathbb{R}^{3\times 3}$ the inertia matrix, $\mathbf{v}_{i}\in\mathbb{R}^{3}$ the velocity, $\mathbf{f}_{i}\in\mathbb{R}^{3}$ the external force, and $\bm{\tau}_{i}\in\mathbb{R}^{3}$ the external torque.
\subsubsection{Euler-Lagrange equation}
The Euler-Lagrange equation of a rigid body system can be formulated as 
\begin{align}\label{EL}
	\mathbf{M}_{i}(\mathbf{q}_{i}) \ddot{\mathbf{q}}_{i}+ \mathbf{C}_{i}(\dot{\mathbf{q}}_{i}, \mathbf{q}_{i}) \dot{\mathbf{q}}_{i}+\mathbf{G}_{i}(\mathbf{q}_{i})=\boldsymbol{\Gamma}_{i}, 
\end{align}
where $\mathbf{q}_{i} \in \mathbb{R}^n$ is the generalized coordinate, $\bm{\Gamma}_{i} \in \mathbb{R}^n$ the control torque, $\mathbf{M}_{i}(\mathbf{q}_{i}) \in \mathbb{R}^{n \times n}$ the inertia matrix, $\mathbf{C}_{i}(\dot{\mathbf{q}}_{i}, \mathbf{q}_{i}) \dot{\mathbf{q}}_{i} \in \mathbb{R}^n$ the centrifugal/Coriolis force vector, and $\mathbf{G}_{i}(\mathbf{q}_{i}) \in \mathbb{R}^n$ the vector of gravitational force.

It should be noted that the Euler-Lagrange system has three important properties as follows:
\begin{itemize}
	\item P.1: The matrix $\dot{\mathbf{M}}_{i}(\mathbf{q}_{i})-2 \mathbf{C}_{i}(\dot{\mathbf{q}}_{i}, \mathbf{q}_{i})$ is skew symmetric.
	\item P.2: The inertia matrix $\mathbf{M}_{i}(\mathbf{q}_{i})$ is a symmetry positive definite matrix, and the norm of Centrifugal/Coriolis matrix satisfies $\|\mathbf{C}_{i}(\dot{\mathbf{q}}_{i}, \mathbf{q}_{i})\| \leq k_{\mathrm{C}_{i}}\|\dot{\mathbf{q}_{i}}\|$, where $k_{\mathrm{C}}$ is a positive real number. The norm of the gravity vector satisfies $\|\mathbf{G}_{i}(\mathbf{q}_{i})\| \leq$ $k_g$, where $k_g$ is a positive real number.
	\item P.3: For vectors $\mathbf{x}, \mathbf{y} \in \mathbb{R}^n$, there exists a linear regressor matrix $\mathbf{Y}_{i}(\dot{\mathbf{q}}_{i}, \mathbf{q}_{i}, \mathbf{x}, \mathbf{y})$ satisfying $\mathbf{M}_{i}(\mathbf{q}_{i}) \mathbf{x}+\mathbf{C}_{i}(\dot{\mathbf{q}}_{i}, \mathbf{q}_{i}) \mathbf{y}+$ $\mathbf{G}_{i}(\mathbf{q}_{i})=\mathbf{Y}_{i}(\dot{\mathbf{q}}_{i}, \mathbf{q}_{i}, \mathbf{x}, \mathbf{y}) \boldsymbol{\Theta}_{i}$, 
	where $\boldsymbol{\Theta}_{i} \in \mathbb{R}^m$ is a constant vector representing the real constant parameter of Euler-Lagrange systems.
\end{itemize}
In fact, the Euler-Lagrange equation is equivalent to Newton-Euler equation. The difference is that the Euler-Lagrange equation is derived based on the principle of the least action, which is a more general and fundamental principle \cite{2001_mechanics}. The Newton-Euler equation is based on Newton's laws of motion. It provides a more direct way of calculating the force and torques acting on a rigid body, which is commonly used in a relatively simple motion control design of rigid bodies. Euler-Lagrange equations provide a complete description of the motion of a rigid body. Thus, it is useful in describing a complex dynamic model when the forces and torques are particularly complicated. In addition, it can design the control input of rotational and translational motion in a unified manner \cite{2002_JGCD_Hadaegh}. 
\subsection{Definitions}
Before introducing the main results, we firstly give some basic definitions of attitude synchronization convergence of multiple rigid body systems. 
Let the consensus set $\mathcal{A} \subseteq \mathbb{SO}(3)^{N}$ be defined as
$
\mathcal{A}=\{\mathcal{R}=\{\mathcal{R}_1,...,\mathcal{R}_N
\}: \mathcal{R}_{i}=\mathcal{R}_{j}, \forall \;i, j=1,...,N \}.
$
A local attitude synchronization definition is firstly given as follows. 
\begin{Def}$^{\cite{2020_Auto_X.Jin}}$
	Consider a multiple rigid body system that consists of $N$ rigid bodies.
	Assuming that the initial attitude $\mathcal{R}_{i}(t_0)$ of each rigid  body is contained in a {\color{black}positively invariant set $\mathcal{B}_{r}(\mathbf{I}_{3})$ where $r<\pi$}, 
	the $\mathit{local \; attitude \; consensus}$ is achieved if $\mathcal{R}(t) \rightarrow \mathcal{A}$ as $t \rightarrow \infty$ for $\;i=1,...,N,\; t\geq 0$.
\end{Def}
According to the property that the set $\mathbb{SO}(3)-\mathcal{B}_{\pi}(\mathbf{I}_{3})$ has measure zero, we have the following almost global attitude synchronization definition.
\begin{Def}$^{\cite{2020_Auto_X.Jin}}$
	Consider a multiple rigid body system that consists of $N$ rigid bodies.
	Assuming that the initial attitude $\mathcal{R}_{i}(t_0)$ of each rigid body is contained in a {\color{black}positively invariant set $\mathcal{B}_{\pi}(\mathbf{I}_{3})$}, 
	the $\mathit{almost \; global \; attitude \; consensus}$ is achieved if $\mathcal{R}(t) \rightarrow \mathcal{A}$ as $t \rightarrow \infty$ for $\;i=1,...,N, t\geq 0$.
\end{Def}

The convergence speed is an important performance in the engineering application of multiple rigid body systems.  Next, we give the finite-time, fixed-time, and prescribed-time synchronization definitions, respectively. 
\begin{Def}$^{\cite{2022_TAC_X.Jin}}$\label{finite-time attitude}
		Consider a multiple rigid body system that consists of $N$ rigid bodies.
	The finite-time attitude synchronization is locally achieved if 
	the attitude $\mathcal{R}_{i}(t, \mathcal{R}_{i}(t_{0}), t_{0}),\forall t_{0}>0, 
	\mathcal{{R}}(t_{0})\in\mathcal{B}_{r}(\mathbf{I}_{3}),\; r<\pi$ reaches $\mathcal{A}$
	 in a settling time $T(	\mathcal{{R}}(t_{0}), t_{0})>0$, i.e., $\text{dist}(	\mathcal{{R}}(T(	\mathcal{{R}}(t_{0}), t_{0}),\mathcal{{R}}(t_{0}),t_0), \mathcal{A})=0$ and $\text{dist}(\mathcal{{R}}(t,\mathcal{{R}}(t_{0}),t_0),\mathcal{A})\equiv0$ for all $t>T(\mathcal{{R}}(t_{0}), t_{0})$. 
\end{Def} 	
\begin{Def}$^{\cite{2022_TAC_X.Jin}}$\label{fixed-time attitude}
		Consider a multiple rigid body system that consists of $N$ rigid bodies.
The fixed-time attitude synchronization is locally achieved if 
	the finite-time attitude synchronization is achieved, and the settling time $T(\mathcal{{R}}(t_{0}), t_{0})$ is uniformly bounded with the initial value $\mathcal{{R}}(t_{0})$, i.e., $\exists T_{\max}>0$ such that $T_{\max}\geq T(\mathcal{{R}}(t_{0}), t_{0}), \;\forall \mathcal{{R}}(t_{0})\in \mathcal{B}_{r}(\mathbf{I}_{3}), r<\pi$.
\end{Def}
\begin{Def}\label{prescribed_time_consensus}
		Consider a multiple rigid body system that consists of $N$ rigid bodies.
	The predefined-time attitude synchronization is locally achieved if 
		the finite-time attitude synchronization is achieved, and the settling time $T$ is predefined such that $T\leq T_{\max}$, where $T_{\max}$ is bounded with initial value $\mathcal{{R}}(t_{0})$, i.e., $\exists T_{\max}>0$ such that $T_{\max}\geq T(\mathcal{{R}}(t_{0}), t_{0}), \;\forall \mathcal{R}({t_0})\in  \mathcal{B}_{r}(\mathbf{I}_{3}), r<\pi$.
\end{Def}
Similarly, the almost global finite-time/fixed-time/prescribed-time attitude synchronization are achieved when $r=\pi$ in Definitions \ref{finite-time attitude}, \ref{fixed-time attitude}, and \ref{prescribed_time_consensus} are satisfied.

\section{Attitude synchronization of multiple rigid body systems}
The attitude synchronization literature can be divided into two categories. One is to use parameterized attitude representations. The attitude of each rigid body evolved on $\mathbb{R}^{3}$ or $\mathbb{S}^{3}$. The other one is to view the attitude as an element on $\mathbb{SO}(3)$. At last, the recent result on networked attitude synchronization is discussed. A literature summary of this section is shown in Table \ref{review1}.
\begin{table*}[htb]
	\begin{tabular}{|c|c|c|c|l|l|c|}
		\hline
		Attitude representations           & Configuration space                       & Measurement               & Convergence                        & Topology     & Communication   & Reference             \\ \hline
		Euler angles                       & $\mathbb{R}^{3}$                  & Absolute                  & Local                              & Fixed     & Continuous      &   \cite{2013_TIE_EULER,2019_TAES_J.Shan}                          \\ \hline
		Modified Rodrigues Parameters      & $\mathbb{R}^{3}$                  & Absolute                  & Local                              & Fixed     & Continuous      &  \cite{2009_SCL_Dimos,2010_TCST_W.Ren_AttitudeSynchronization,2010_Auto_Z.Meng_attitude}                     \\ \hline
		\multirow{4}{*}{Axis-angles}       & \multirow{4}{*}{$\mathbb{R}^{3}$} & \multirow{2}{*}{Absolute} & \multirow{2}{*}{Almost global}     & Switching & Continuous      &        \cite{2014_Auto_J.T_attitude}                          \\ \cline{5-7} 
		&                                           &                           &                                    & Fixed     & Event-triggered &  { \cite{2020_Auto_X.Jin} } \\ \cline{3-7} 
		&                                           & \multirow{2}{*}{Relative} & \multirow{2}{*}{Local}             & Switching & Continuous      &   \cite{2016_Auto_J.Thunberg}                    \\ \cline{5-7} 
		&                                           &                           &                                    & Switching & Event-triggered &  \cite{2022_Auto_X.Jin_Event}               \\ \hline
		\multirow{5}{*}{Quaternions}       & \multirow{5}{*}{$\mathbb{S}^{3}$} & \multirow{3}{*}{Absolute} & \multirow{3}{*}{Global(Unwinding)} & Fixed     & Continuous      &    \cite{2018_Auto_J.Huang,2020_TAC_Dimos_attitude}                       \\ \cline{5-7} 
		&                                           &                           &                                    & Fixed     & Sampled-data    & \cite{2022_Auto_J.Huang}   \\ \cline{5-7} 
		&                                           &                           &                                    & Switching & Continuous      & \cite{2018_Auto_J.Huang,2020_TAC_Dimos_attitude}  \\ \cline{3-7} 
		&                                           & \multirow{2}{*}{Absolute} & \multirow{2}{*}{Global}            & Fixed     & Continuous      &   \cite{2012_TAC_A.R.Teel,2018_Auto_H.Gui}                    \\ \cline{5-7} 
		&                                           &                           &                                    & Fixed     & Event-triggered & \cite{2019_Tmech_D.Zhang,2022_TCS1_D.Zhang} \\ \hline
		\multirow{4}{*}{Rotation matrices} & \multirow{4}{*}{$\mathbb{SO}(3)$}                    & Absolute                  & Local                              & Fixed     & Continuous      &     \cite{2018_TMech_Y.Zou,2019_TCYB_Z.Meng}                      \\ \cline{3-7} 
		&                                           & Absolute                  & Almost global                      & Fixed     & Continuous      & \cite{2009_Auto_Sarlette}                           \\ \cline{3-7} 
		&                                           & Relative                  & Local                              & Fixed     & Continuous      &        \cite{2013_TAC_Tron}                  \\ \cline{3-7} 
		&                                           & Relative                  & Almost global                      & Fixed     & Continuous      &      \cite{2018_Auto_JT_Dynamic,2018_TAC_JM_n-sphere,2020_Auto_J.Thunberg}                   \\ \hline
	\end{tabular}
	\caption{A literature summary of coordination control of multiple rigid body systems.}
	\label{review1}
\end{table*}
\subsection{Parameterized attitude synchronization on $\mathbb{R}^{3}$}
An early work of a cooperative control of multiple rigid bodies by using parameterized attitude synchronizations is considered in the leader-follower framework \cite{2009_SCL_Dimos}. The attitude of rigid bodies is represented by Modified Rodriguez parameters (MRPs). The kinematics of rigid body systems by using MRPs are shown in (\ref{MRP}).
The control objectives in this work are two aspects. One is that 
the followers are "dragged" by leaders into the convex hull of the leaders' orientations. 
The leaders are assumed to converge to the final orientations and the angular velocities will converge to zero, i.e.,
\begin{align}
	\bm{\sigma}_i=\bm{\sigma}_i^d, \quad \bm{\omega}_i=0, \quad i \in \mathcal{N}^l,
\end{align}
where $\mathcal{N}^{l}$ is the node set of leaders.
The other case is to drive the leaders to the desired relative orientations. 
The relative orientation for each pair of leaders $(i,j)\in \mathcal{E}$ could be different which is given as $\bm{\sigma}_{ij}^{d}\in \mathbb{R}^{3}$.
The problem is motivated by the real applications in the multiple satellite scheme. The coordination observation task requires a group of satellite covering a specific area. The leaders' orientations dictate the "boundary" of the area to be covered \cite{2010_TCST_W.Ren_AttitudeSynchronization}.

For the first case, the control law of the followers is given as 
\begin{align}\label{Rod_Protocol1}
	\bm{\tau}_i=-\bm{G}^{\top}\left(\bm{\sigma}_i\right) \sum_{j \in \mathcal{N}_i}\left(\bm{\sigma}_i-\bm{\sigma}_j\right)-\sum_{j \in \mathcal{N}_i}\left(\bm{\omega}_i-\bm{\omega}_j\right), \quad i \in \mathcal{N}^f,
\end{align}
where $\mathcal{N}^{f}$ is the node set of followers.
For the second case, the control law of the leaders is given as 
\begin{align}\label{Rod_Protocol2}
	& \bm{\tau}_i=-\bm{G}^{\top}\left(\bm{\sigma}_i\right) \sum_{j \in \mathcal{N}_i^l}\left(\bm{\sigma}_i-\bm{\sigma}_j-\bm{\sigma}_{i j}^d\right)-\sum_{j \in \mathcal{N}_i^l}\left(\bm{\omega}_i-\bm{\omega}_j\right),\nonumber\\
	 &\quad i \in \mathcal{N}^l \backslash\{\alpha\},
\end{align}
where a leader $\alpha\in\mathcal{N}^{l}$ is a reference attitude with respect to the desired relative orientation. Here, it is assumed that this leader has already been stabilized to the desired orientation, i.e., $\bm{\sigma}_\alpha=\bm{\sigma}_\alpha^d, \quad \bm{\omega}_\alpha=0$.

The form of protocols (\ref{Rod_Protocol1}), (\ref{Rod_Protocol2}) are similar to the second-order consensus protocol for multi-agent systems. However, the difference is that the attitude is governed by the dynamic model (\ref{dynamics_newton}). Based on the connected topology condition and the properties of the matrix $\bm{G}(\bm{\sigma}_i)$, we can show that the protocol (\ref{Rod_Protocol1}) can drive the followers to the convex hull of the leaders’
orientations. Furthermore, if no global objective is imposed by the leaders, i.e., $\mathcal{N}^l=\varnothing$. The following proposed distributed attitude synchronization protocol can drive the group of rigid bodies to a common constant orientation with zero angular velocities,
\begin{align}
	\bm{\tau}_i=-\bm{G}^{\top}\left(\bm{\sigma}_i\right) \sum_{j \in \mathcal{N}_i}\left(\bm{\sigma}_i-\bm{\sigma}_j\right)-\sum_{j \in \mathcal{N}_i}\left(\bm{\omega}_i-\bm{\omega}_j\right)-a_i \bm{\omega}_i,
\end{align}
where $a_i$ is the gain of the damp term.

The above protocols require the angular velocity measurement. An angular velocity-free framework motivated by the passivity approach is proposed for multiple rigid body systems \cite{2010_TCST_W.Ren_AttitudeSynchronization}. The control protocol is designed as follows,
\begin{align}
	& \dot{\hat{\bm{x}}}_i=\bm{\varGamma} \hat{\bm{x}}_i+\sum_{j=1}^n b_{i j}\left(\bm{\sigma}_i-\bm{\sigma}_j\right)+\kappa \bm{\sigma}_i \label{estimator}\\
	& \bm{y}_i=\bm{P} \bm{\varGamma} \hat{\bm{x}}_i+P \sum_{j=1}^n b_{i j}\left(\bm{\sigma}_i-\bm{\sigma}_j\right)+\kappa \bm{P}\bm{\sigma}_i \\
	& \tau_i=-\bm{G}^{\top}\left(\bm{\sigma}_i\right)\left[\sum_{j=1}^n a_{i j}\left(\bm{\sigma}_i-\bm{\sigma}_j\right)+a_{i(n+1)}\left(\bm{\sigma}_i-\bm{\sigma}_c^r\right)+\bm{y}_i\right],\label{ren}
\end{align}
where $\bm{\sigma}_{c}^{r}\in \mathbb{R}^{3}$ denotes the constant reference attitude for each rigid body, $a_{ij}$ and $b_{ij}$ are entries of two adjacency matrices $\mathbf{A}$ and $\mathbf{B}$ associated with the communication graph, $\kappa$ is a positive parameter, and $\bm{P}=\bm{P}^{\top}\in \mathbb{R}^{3 \times 3}$ is the solution of the Lyapunov equation $\bm{\varGamma}^{\top}\bm{P}+\bm{P}\bm{\varGamma}=-\bm{Q}$ with $\bm{Q}=\bm{Q}^{\top}>0\in \mathbb{R}^{3 \times 3}$.
Note that the term $\sum_{j=1}^n b_{i j}\left(\bm{\sigma}_i-\bm{\sigma}_j\right)$ in (\ref{estimator}) provides the relative damping between neighboring rigid bodies,  and the output signal $\bm{y}_{i}$ replaces the angular velocity feedback in the control torque.

{\color{black}Note that MRPs describe the attitude as a vector in Euclidean spaces, 
which makes the synchronization protocol design and convergence analysis convenient.
There are other results of attitude synchronization based on MRPs \cite{2022_TNSE_X.Jin}.
Motivated by the precise requirement of completing time in some aerospace tasks, finite-time attitude synchronization has been widely studied \cite{2010_Auto_Z.Meng_attitude}.
A distributed finite-time attitude containment control is studied for multiple rigid body systems \cite{2010_Auto_Z.Meng_attitude}. The multiple stationary leaders and dynamic leaders are both considered. Two kinds of distributed protocols are designed to guarantee that followers' attitudes converge to the convex hull formed by leaders in finite time.
However, the estimation settling time of the finite-time consensus protocol is quite conservative. In addition, it depends on the initial attitude and parameters of rigid bodies. 
A prescribed-time attitude consensus problem is studied using MRPs where the users can predetermine the settling time \cite{2022_TASE_X.Chuang}. 
However, the parameterized attitude representation on $\mathbb{R}^{3}$ has a singularity problem. For MRPs, the singularity point corresponds to the attitude that the rotation angle approaches $2\pi$} \cite{2010_TCST_W.Ren_AttitudeSynchronization}. 
\subsection{Parameterized attitude synchronization on $\mathbb{S}^{3}$}
The unit quaternion is a global attitude representation \cite{1993_JAS_M.Shuster_attitude}.
A coordinated attitude control problem for multiple rigid body systems is investigated with communication delays and without angular velocity measurements based on unit quaternions \cite{2012_TAC_Abdessameud_Attitude+Time-delay}. This work proposed a virtual dynamic system approach to handle the communication delay and remove the requirement of angular velocity measurements. 
The first virtual system associates to each rigid body is formulated as 
\begin{align}\label{virtual_state1}
	\dot{\mathbf{Q}}_{v_i}=\frac{1}{2} \mathbf{T}\left(\mathbf{Q}_{v_i}\right) \boldsymbol{\omega}_{v_i}
\end{align} 
for $i \in \mathcal{N}$, where $\mathbf{Q}_{v_i}=\left(\mathbf{q}_{v_i}^{\top}, \eta_{v_i}\right)^{\top}$ is the unit quaternion representing the state of the virtual system (\ref{virtual_state1}). The  $\mathbf{Q}_{v_i}(0)$ can be initialized arbitrarily, and $\omega_{v_i}$ is the virtual angular velocity input which will be designed later. 
The matrix $\mathbf{T}\left(\mathbf{Q}_{v_i}\right)$ is defined as: $\mathbf{T}\left(\mathbf{Q}_{v_i}\right)=\left(\begin{array}{c}\eta_{v_i} \mathbf{I}_3+\mathbf{S}\left(\mathbf{q}_{v_i}\right) \\ -\mathbf{q}_{v_i}^{\top}\end{array}\right)$.

The second virtual system is formulated as 
\begin{align}\label{virtual_state2}
	\dot{\mathbf{P}}_{i}=\frac{1}{2} \mathbf{T}\left(\mathbf{P}_{i}\right) \boldsymbol{\beta}_{i},
\end{align} 
where $\mathbf{P}_{i}$ is the unit quaternion representing the state of the virtual system (\ref{virtual_state2}), and the initial values can be given arbitrarily. $\mathbf{T}\left(\mathbf{P}_{i}\right)$ is given similar to (\ref{virtual_state1}), and $\beta_{i}\in \mathbb{R}^{3}$ is an input to be determined. 

The main idea in this approach is to design the control input $\bm{\tau}_i$ for each rigid body as well as the input for virtual systems associated with each rigid body. The control input $\bm{\tau}_i$ is designed based on the signal constructed by the state of these two virtual systems without requiring the angular velocity measurement. The control input for each rigid body system is designed as 
\begin{align}\label{control_input}
	\boldsymbol{\tau}_i=\mathbf{H}_i\left(\dot{\boldsymbol{\omega}}_{v_i}, \boldsymbol{\omega}_{v_i}, \mathbf{Q}_i^e\right)-k_i^p \mathbf{q}_i^e-k_i^d \tilde{\mathbf{q}}_i^e, \; i\in \mathcal{N},
\end{align} 
where
\begin{align}
	\mathbf{H}_i(\dot{\boldsymbol{\omega}}_{v_i} \boldsymbol{\omega}_{v_i}, \mathbf{Q}_i^e)=&\;\mathbf{I}_{f_i} \mathbf{R}\left(\mathbf{Q}_i^e\right) \dot{\boldsymbol{\omega}}_{v_i}+\left(\mathbf{R}\left(\mathbf{Q}_i^e\right) \boldsymbol{\omega}_{v_i}\right)^{\wedge} \mathbf{I}_{f_i} \mathbf{R}\left(\mathbf{Q}_i^e\right) \boldsymbol{\omega}_{v_i},
\end{align} 
and $k_i^p$, $k_i^d$ are strictly positive scalar gains, $\mathbf{q}_i^e$ is the vector part of the
unit quaternion $\mathbf{Q}_{i}^{e}=\left(\mathbf{q}_i^{e^{\top}}, \eta_i^e\right)^{\top}$ which is defined as 
$\mathbf{Q}_i^e=\mathbf{Q}_{v_i}^{-1} \odot \mathbf{Q}_i$, and $\widetilde{\mathbf{q}}_i^e$ is the vector part of the
unit quaternion $\widetilde{\mathbf{Q}}_{i}^{e}=\left(\widetilde{\mathbf{q}}_i^{e^{\top}}, \widetilde{\eta}_i^e\right)^{\top}$ which is defined as 
$\widetilde{\mathbf{Q}}_i^e=\mathbf{P}_{i}^{-1} \odot \mathbf{Q}_{i}^{e}$. The control inputs for two virtual systems are given as 
\begin{align}\label{v_input_1}
\dot{\boldsymbol{\omega}}_{v_i}=-k_i^{{\omega}} \boldsymbol{\omega}_{v_i}-\sum_{j=1}^n k_{i j} \overline{\mathbf{q}}_{v_{i j}},
\end{align}
and
\begin{align}\label{v_input_2}
\mathbf{\beta}_{i}=\lambda_{i}\widetilde{\mathbf{q}}_{i}^{e}, \; i\in \mathcal{N},
\end{align}
where $\omega_{v_i}(0)$ can be selected arbitrarily, $\overline{\mathbf{q}}_{v_{i j}}$ is the vector part of the unit quaternion $\overline{\mathbf{Q}}_{v_{i j}}=\mathbf{Q}_{v_j}^{-1}\left(t-\tau_{i j}\right) \odot \mathbf{Q}_{v_i}$ 
The scalar gains $k_l^q>0, k_i^\omega>0, \lambda_{i}>0$ for $i \in \mathcal{N}$, and $k_{i j} \geq 0$ is the $(i, j)^{t h}$ entry of the adjacency matrix of the weighted undirected graph $\mathcal{G}$. 
Based on the control input (\ref{control_input})-(\ref{v_input_2}), if the control gain satisfies $k_{i}^{w}-\sum_{j=1}^{N}(\frac{k_{ij}}{4})(\epsilon+(\frac{\tau^{2}}{\epsilon}))>0$ where $\tau$ is the upper bound of the time-varying communication delays such that $\tau_{i j}\leq \tau$ for $(i,j)\in\mathcal{E}$, the attitude synchronization can be attained. 

{\color{black}
The difference between the angular velocity-free approach (\ref{control_input}) and (\ref{ren}) lie in the dynamics of attitude. The attitude configuration space is Euclidean space in (\ref{estimator}). However, in (\ref{virtual_state1}), the virtual state is governed by the unit quaternion dynamics, which is nonlinear. The benefit of the approach in (\ref{control_input}) is that it can be used in the relative attitude measurement case, and the unit quaternion can describe attitude globally without singularity problems.
Numerous results on attitude synchronization based on unit quaternions have been made. An angular velocity-free leader-follower attitude consensus with a dynamic leader is solved by proposing a distributed unit quaternion-based attitude feedback control law \cite{2016_Auto_J.Huang_Attitude}. A distributed observer is proposed to estimate the leader's state, and an auxiliary system is designed to compensate for the angular velocity. 
Following this distributed observer approach, the leader-following attitude consensus that is subject to jointly connected switching topologies \cite{2018_Auto_J.Huang} and sampled-data scheme \cite{2022_Auto_J.Huang} are studied by using unit-quaternion representations. 

Although the unit quaternion can describe attitudes globally, it is a non-unique representation. The non-unique attitude representation can lead to an undesirable phenomenon called unwinding.
In unwinding, for certain initial conditions under attitude kinematic (\ref{quaternion_kinematic}), the trajectories can undergo a homoclinic-like orbit that starts close to the desired attitude equilibrium \cite{2011_ControlSystemMagazine}.
Thus, the quaternion-based attitude synchronization scheme may achieve global synchronization. However, the synchronization state can be stable or unstable \cite{2012_TAC_A.R.Teel}. } Motivated by this fact, a hybrid feedback using unit quaternions that achieves the global attitude synchronization is proposed for each rigid body \cite{2012_TAC_A.R.Teel}.
The unwinding phenomenon can be avoided by using a logic variable associated with each pair of rigid bodies which determines the sign of a torque input component. 

Let
$
h=\{h_{1},h_{2},...,h_{M}\}\in \{-1,1\}^{M}
$
denote a binary logic variable vector, where $h_{k}$ is associated with each link $k \in \mathcal{M}$ in the graph. 
Let the flow set $C_{i}$ and jump set $D_{i}$ for rigid body $i$ be given as 
\begin{align}
	C_i & =\left\{x \in \mathcal{X}: \forall k \in \mathcal{M}_i^{+}, h_k \tilde{\eta}_k \geq-\delta\right\} \\
	D_i & =\left\{x \in \mathcal{X}: \exists k \in \mathcal{M}_i^{+}, h_k \tilde{\eta}_k \leq-\delta\right\},
\end{align}
where $\delta$ is a positive constant, $\mathcal{X}$ is the state space, and $\tilde{\eta}_{k}$ is the scalar component of the relative attitude error for each link $k \in \mathcal{M}$. 
The hybrid dynamics of the binary logic variable $h_{k}$ is given as 
\begin{align}
	\begin{array}{lll}
		\forall k \in \mathcal{M}_i^{+} & \dot{h}_k=0 & x \in C_i \\
		& h_k^{+} \in h_k \overline{\operatorname{sgn}}\left(h_k \tilde{\eta}_k+\alpha\right) & x \in D_i,
	\end{array}
\end{align}
where $\alpha \in [0,\delta)$, and the set-valued map  $\overline{\operatorname{sgn}}: \mathbb{R} \rightrightarrows\{-1,1\}$ is defined as
$$
\overline{\operatorname{sgn}}(s)= \begin{cases}s /|s| & s \neq 0 \\ \{-1,1\} & s=0 .\end{cases}
$$
Based on the logic variable and the reference angular velocity signal $\omega_d$, the control torque is given as, 
\begin{align}\label{hybrid_input}
	\bm{\tau}_i=-\left(\mathbf{J}_i \bm{\omega}_i\right)^{\wedge} \bm{\omega}_d-\sum_{k=1}^M b_{i k} h_k \ell_k \tilde{\epsilon}_k-K_i \bar{\bm{\omega}}_i,
\end{align}
where $l_{k}>0, \; \forall k \in M$, $\bar{\bm{\omega}}_i=\bm{\omega}_i-\bm{\omega}_d$, and $K_{i}=K_{i}^{\top}>0,\; \forall i\in \mathcal{N}$. 

The binary logic variable incorporated in the control law (\ref{hybrid_input}) can hysterically switch the sign of a torque component which has an anti-unwinding property.  
In addition, the hybrid control law (\ref{hybrid_input}) can achieve the robust attitude synchronization under the connected acyclic graphs, and manage a trade-off between unwinding and robustness by adjusting the hysteresis width $\delta$.
Based on the hybrid control, there are fruitful results on global attitude synchronization \cite{2022_TCS1_D.Zhang}. 
For example, the global finite-time attitude consensus is investigated with quaternion-based hybrid controllers \cite{2018_Auto_H.Gui}. 
A hybrid attitude tracking control is studied based on the event-triggered mechanism \cite{2019_TMecha_Y.Tang}.

The rotation matrix is a global and unique attitude representation method. 
However, the closed-loop dynamics by using a continuous state-feedback based on rotation matrices usually has undesired equilibrium points which are unstable.
The fundamental difficulty is the underlying space of rotation matrices is a Lie group, which is not homeomorphic to $\mathbb{R}^{n}$. 
Inspired by the above hybrid method \cite{2012_TAC_A.R.Teel}, a hybrid-based attitude tracking controller on $\mathbb{SO}(3)$ is proposed to obtain the global result \cite{2017_Auto_Tayebi}.
Following the idea, an angular velocity-free global attitude tracking on $\mathbb{SO}(3)$ and $\mathbb{SE}(3)$ are further studied, respectively
\cite{2022_TAC_Tayebi, 2018_TAC_Tayebi}.

\subsection{Attitude synchronization on $\mathbb{SO}(3)$}
Due to the topological complexities of $\mathbb{SO}(3)$, there is no smooth state-feedback control that can globally solve the attitude stabilization. Thus, the best result of using the smooth control protocol is almost global attitude synchronization \cite{2018_TAC_JM_n-sphere}. 
The attitude synchronization is considered with switching topologies for multiple rigid body systems \cite{2014_Auto_J.T_attitude}. 
The rotation of each rigid body is described by the axis-angle representation which can almost globally represent the attitude. 
The axis-angle representation of the absolute attitude measurement and the relative attitude measurement can be calculated by using the logarithm map
\begin{align}
{\mathbf{x}}_{i}^{\wedge}=\log(\mathcal{R}_{i}),
\end{align}
and 
\begin{align}
{\mathbf{x}}_{ij}^{\wedge}=\log(\mathcal{R}_{i}^{\top}\mathcal{R}_{j}),
\end{align}
where ${\mathbf{x}}_{i}^{\wedge}\in \mathfrak{so}(3)$ is the skew-symmetric matrix generated by $\mathbf{x}_{i}=[x_{i}^{1},x_{i}^{2},x_{i}^{3}]\in \mathbb{R}^{3}$.
Based on the absolute attitude measurement and the relative attitude measurement information, the attitude synchronization protocols are given as follows,
\begin{align}\label{absolute}
	\bm{\omega}_{i}^{a}=\sum_{j \in \mathcal{N}_i(t)}a_{ij}(t)(\mathbf{x}_{j}-\mathbf{x}_{i}),
\end{align}
and 
\begin{align}\label{relative}
	\bm{\omega}_{i}^{r}=\sum_{j \in \mathcal{N}_i(t)}a_{ij}(t)(\mathbf{x}_{ij}),
\end{align}
where $a_{ij}(t)$ is a weighted matrix associated with the time-varying graph $\mathcal{G}(t)$ in Definition \ref{switching},  $\bm{\omega}_{i}^{a}$ and $\bm{\omega}_{i}^{r}$ are the angular velocity inputs based on absolute attitude measurements and relative attitude measurements for rigid body $i$, respectively. 

It can be proven that the first protocol (\ref{absolute}) guarantees the positive invariance of the open ball $\mathcal{B}_{\pi}(\mathbf{0})$ which can almost globally cover $\mathbb{SO}(3)$. Thus, the almost global attitude synchronization is achieved by using (\ref{absolute}). 
For the relative attitude measurement only case, the convergence result is based on the convex property on the local set $\mathcal{B}_{\frac{\pi}{2}}(\mathcal{Q})$, where $\mathcal{Q}$ is an arbitrary rotation on $\mathbb{SO}(3)$.  
In addition, the protocol input (\ref{relative}) can be interpreted from the geometric view, which is inward-pointing to the boundary of the convex hull on $\mathcal{B}_{\frac{\pi}{2}}(\mathcal{Q})$. It can be shown that the convex hull is shrinking and further shrinks to one point, which achieves the local attitude synchronization \cite{2014_AMS_B.Afsari.}. 
The above results \cite{2014_Auto_J.T_attitude} are quite interesting since it only uses the well-known consensus protocol as shown in (\ref{absolute}) to achieve the attitude synchronization, which allows methods that are suitable to the Euclidean space $\mathbb{R}^{3}$. 
The local and almost global finite-time attitude consensus in Definition \ref{finite-time attitude} are achieved based on a discontinuous attitude consensus protocol \cite{2018_TAC_J.Wei_Finite-time}.  
However, the discontinuous control input signal may not be appropriate for the implementation in the mechanical system, which is harmful to the actuator. 
A fixed-time attitude consensus protocol is designed by constructing a class of particularly continuous functions \cite{2022_TAC_X.Jin}.

Different from the complete attitude synchronization on $\mathbb{SO}(3)$, a reduced attitude can be considered as an element on two-dimensional spheres \cite{2011_ControlSystemMagazine}.
Incomplete attitude synchronization corresponds to practical problems such as moving along a common direction in flocks and pointing to a common direction in a network of satellites. 
A common framework of synchronization of agents on $\mathbb{S}^{2}$ and $\mathbb{SO}(3)$ is proposed under switching topologies \cite{2020_TAC_Dimos_attitude}. The complete attitude synchronization is cast as synchronization on $\mathbb{S}^{3}$ and the incomplete attitude synchronization is cast as synchronization on $\mathbb{S}^{2}$. 
It should be noted that the consensus problem on $\mathbb{S}^{n}$ has a strong application background, including reduced attitude synchronization \cite{2020_TAC_Dimos_attitude}, self-synchronizing oscillators \cite{1996_PRL,1997_PRL,2020_Chaos_oscillator}, and quantum consensus \cite{2016_TAC_Shi,2017_tac_Shi}. 
The almost global consensus result is established for a class of consensus protocols on $n$-spheres except for the circle in \cite{2018_TAC_JM_n-sphere}. 
The agent's state $\mathbf{x}_{i}\in \mathbb{S}^{n}$ and dynamics is governed by 
\begin{align}\label{dyanmic_sphere}
	\dot{\mathbf{x}}_i=\mathbf{u}_i-\left\langle\mathbf{u}_i, \mathbf{x}_i\right\rangle \mathbf{x}_i=\left(\mathbf{I}-\mathbf{X}_i\right) \mathbf{u}_i=\mathbf{P}_i \mathbf{u}_i,
\end{align}
where $\mathbf{u}_i: \mathcal{I}_i \rightarrow \mathbb{R}^{n+1}$ is the input signal of agent $i, \mathbf{X}_i=$ $\mathbf{x}_i \otimes \mathbf{x}_i$, and $\mathbf{P}_i=\mathbf{I}-\mathbf{X}_i$ for all $i \in \mathcal{V}$.
{\color{black}The dynamics of the state $\mathbf{x}_{i}\in \mathbf{S}^{n}$ (\ref{dyanmic_sphere}) can be derived from the dynamics on $\mathbb{SO}(n)$. In fact, the state $\mathbf{x}_{i}\in \mathbf{S}^{n}$ can be seen as a column of the matrix $\mathcal{R}_{i}\in \mathbb{SO}(n)$. From this point of view, letting $\mathcal{R}_{i}[1,0,..., 0]^{\top}=\mathbf{x}_{i}$, the dynamics of ${\mathbf{x}}_{i}$ can be written as  $\dot{\mathbf{x}}=\mathbf{P}_{i}\mathbf{u}_{i}$, where $\mathbf{P}_{i}$ is a projector that transforms the input $\mathbf{u}_{i}$ onto the tangent space at the point $\mathbf{x}_{i}$ \cite{2018_Auto_Thunberg}.}

The consensus algorithm is derived by taking the gradient of the following Lyapunov function $V:(\mathbb{S}^{n})^{N}\rightarrow [0, +\infty)$, 
\begin{align}
	V\left(\left(\mathbf{x}_i\right)_{i=1}^N\right)=\sum_{\{i, j\} \in \mathcal{E}} \int_0^{s_{i j}} f_{i j}(r) d r,
\end{align}
where $s_{ij}=1-\langle\mathbf{x}_{i},\mathbf{x}_{j}\rangle$ and $f_{ij}: \mathbb{R}\rightarrow \mathbb{R}$ is a real analytic function satisfying the following condition:
i) $f_{i j}>0$;
ii) $f_{i j}=f_{j i}$; and
iii) $\left(n-2+s_{i j}\right) s_{i j} f_{i j}-\left(2-s_{i j}\right) s_{i j}^2 f_{i j}^{\prime}>0$
for all $s_{i j} \in(0,2]$ and all $\{i, j\} \in \mathcal{E}$.
By embedding the sphere $\mathbb{S}^{n}$ in $\mathbb{R}^{n+1}$, the extension function $U: (\mathbb{R}^{n+1})^{N}\rightarrow \mathbb{R}$ can be given by
\begin{align}
	U\left(\left(\mathbf{x}_i\right)_{i=1}^N\right)=\sum_{\{i, j\} \in \mathcal{E}} \int_0^{s_{i j}} f_{i j}(r) d r.
\end{align}  
Then, the control protocol can be obtained by
\begin{align}\label{sphere}
	\mathbf{u}_{i}=-\nabla_i U=-\sum_{j \in \mathcal{N}_i} \frac{d U}{d s_{i j}} \nabla_i s_{i j}=\sum_{j \in \mathcal{N}_i} f_{i j}\left(s_{i j}\right) \mathbf{x}_j,
\end{align}
where $	\nabla_i $ denotes $\nabla_{\mathbf{x}_{i}}$.

It can be shown that when the protocol (\ref{sphere}) is utilized, the almost global consensus on $n$-sphere, $n\in \mathbb{N}\setminus \{1\}$ is reached.  
To show this result, the first step is to prove that the consensus set $\mathcal{C}$ is asymptotically stable. 
This fact can be obtained since the right hand side of (\ref{dyanmic_sphere}) points toward the geodesically convex hull of $\{\mathbf{x}_{j}| j\in \mathcal{N}_{j}\}$ on $\mathbb{S}^{n}$. The second step is to prove the instability of the undesired equilibrium points on $\mathbb{S}^{n}$.
This fact is derived by using the linearized around the equilibrium points and due to the property of the gain function $f_{ij}$.
Then, combining these facts, the almost global consensus results can be obtained. This analysis procedure is also suitable for consensus problems on more general manifolds. 

The attitude of rigid bodies is an element in the special orthogonal group, i.e., $\mathbb{SO}(3)$. More generally, it can be considered as a smooth manifold. A distributed consensus algorithm with the states lying in a Riemannian manifold is proposed for a multi-agent system \cite{2013_TAC_Tron}. 
The idea of the algorithm is to formulate the consensus problem as an optimization problem and define the cost function on Riemannian manifold, which describes the disagreement of distances for multi-agent systems. 
The cost function on Riemannian manifold is given as 
\begin{align}
	\varphi(\mathbf{x})=\frac{1}{2} \sum_{\{i, j\} \in E} d^2\left(\mathbf{x}_i, \mathbf{x}_j\right),
\end{align}
where $\mathbf{x}_{i}\in \mathcal{M}$, $\mathcal{M}$ denotes a Riemannian manifold, and $d(\mathbf{x}_i, \mathbf{x}_j)$ is the geodesic distance between $\mathbf{x}_i$ and $\mathbf{x}_j$. The distributed algorithm of each node is obtained through Riemannian gradient descent. 
The update rule of each agent is obtained by calculating the gradient of $\varphi$ with respect to $\mathbf{x}_{i}$ as follows,
\begin{align}\label{manifold}
	\operatorname{grad}_{\mathbf{x}_i} \varphi=\frac{1}{2} \operatorname{grad}_{\mathbf{x}_i} \sum_{j \in \mathcal{N}_i} d^2\left(\mathbf{x}_i, \mathbf{x}_j\right)=-\sum_{j \in \mathcal{N}_i} \log _{\mathbf{x}_i}\left(\mathbf{x}_j\right),
\end{align}
where $\log$ is the logarithm map.   
The main result discusses relationships between the convergence of the algorithm and domain of attraction on Riemannian manifold as well as topology graphs. 
Based on (\ref{manifold}), if the initial states are contained in the set $\mathcal{S}_{conv}=\{\mathbf{x}\in \mathcal{M}^{N}: \varphi(\mathbf{x})<\frac{(r^{*})}{2D} \}$, where $D$ is the diameter of the graph $\mathcal{G}$ and $r^{*}$ is the convexity radius of the manifold, the consensus can be achieved if step size $\epsilon\in(0, 2\mu_{\max}^{-1})$ is admissible.
The upper bound of the parameter $\mu_{\max}$ for the step size is determined according to the curvature of the manifold.

The above consensus result requires undirected graphs and the convexity property for the manifold in the convergence analysis. 
To relax the requirement,
a novel control scheme for synchronization on $\mathbb{SO}(d)$ is proposed in a distributed manner \cite{2018_Auto_JT_Dynamic}. Based on a QR-factorization approach, a dynamic feedback control algorithm is proposed for synchronization of the $k$ first columns of the matrix on $\mathbb{SO}(d)$. Based on the control scheme, the almost global convergence is achieved under strongly connected graphs. 
A more general result of synchronization on Stiefel manifolds is shown based on the high-dimensional Kuramoto model which covers the case of $\mathbb{S}^{n}$ and $\mathbb{SO}(n)$ \cite{2020_Auto_JT}. 
Inspired by the above approach \cite{2018_TAC_JM_n-sphere}, it is proven that the almost global synchronization of the generalized Kuramoto model on Stiefel manifold $St(p,n)$ is achieved for any connected graphs if the condition $p \leq \frac{2}{3}n-1$ is satisfied \cite{2020_Auto_JT}. 
Furthermore, synchronization on Riemannian manifolds is considered in the sense of geodesic distances and chordal distances for manifolds, respectively \cite{2021_TAC_JM}. 
It is shown that, if the manifold  is multiply connected or contains a closed geodesic that is of locally minimum length in a space of closed curves, the consensus algorithms are multi-stable.
Note that the previous result on $\mathbb{S}^{n}$ and $\mathbb{SO}(n)$ is a special case of this result \cite{2021_TAC_JM}.

{\color{black}\subsection{Sampled-data based attitude synchronization}
In general, attitude synchronization is realized by means of  information sharing through multiple rigid body networks. 
The data in communication networks is transmitted in the form of digital signals based on sampled-data mechanism rather than continuous signals \cite{2022_Auto_J.Huang}. In addition, due to the limited bandwidth, network traffic congestion is unavoidable leading to network-induced delays \cite{2012_TAC_Abdessameud_Attitude+Time-delay}.
Recently, attitude synchronization under networked constraints has been studied in different aspects such as communication time delays \cite{2016_TAC_H.Du}, sampled-data mechanism \cite{2022_Auto_J.Huang}, and event-triggered mechanism \cite{2020_Auto_X.Jin}.
A leader-following consensus of multiple rigid body systems is studied under a sampled-data communication setting \cite{2022_Auto_J.Huang}.
The dynamics of the leader system is governed by the following system
\begin{align}
		\dot{v} & =S v, \quad \Omega_0=E v \\
		\dot{\bm{q}}_0 & =\frac{1}{2} \bm{q}_0 \odot 
		\left[\begin{array}{l}
		0 \\
			\Omega_0
		\end{array}\right],
\end{align}
where $S \in \mathbb{R}^{m \times m}$ and $E \in \mathbb{R}^{3 \times m}$ are constant matrices with the $\operatorname{pair}(E, S)$ detectable, $v \in \mathbb{R}^m, \Omega_0 \in \mathbb{R}^3$, and $\bm{q}_0=\operatorname{col}\left(\bar{\bm{q}}_0,\hat{\bm{q}}_0\right) \in \mathbb{S}^{3}$, $\hat{\bm{q}}_0\in \mathbb{R}^{3}, \bar{\bm{q}}_0\in \mathbb{R}$.

A sampled-data distributed observer is proposed to estimate the state of the leader system as follows, when $t\in [t_{s},t_{s+1})$,
	\begin{align}
		& \dot{\xi}_i(t)=S \xi_i(t)+L \sum_{j=0}^N a_{i j} E\left(\xi_j\left(t_s\right)-\xi_i\left(t_s\right)\right), \\
		& \dot{\eta}_i(t)=\frac{1}{2} \eta_i(t) \odot \mathbf{Q}\left(\zeta_i(t)\right)+\mu \sum_{j=0}^N a_{i j}\left(\eta_j\left(t_s\right)-\eta_i\left(t_s\right)\right),
	\end{align}
	where $L\in\mathbb{R}^{m \times 3}$ is a positive definite matrix, $\mu$ is a positive number, $\xi_{0}=v$, $\eta_{0}=\bm{q}_{0}$, and $\zeta_i=E\xi_{i}$. $t_{s+1}=t_{s}+T_{s}, s\in \mathbb{N}$ are the sampling instants, and $T_{s}\in [\underline{h},\overline{h}]$. 
	One of the main results is to determine the explicit upper bound  for the sampling intervals to guarantee
	the validity of the sampled-data distributed observer.}

\begin{figure}[h]	
	\centering
	{
		\includegraphics[scale=0.3]{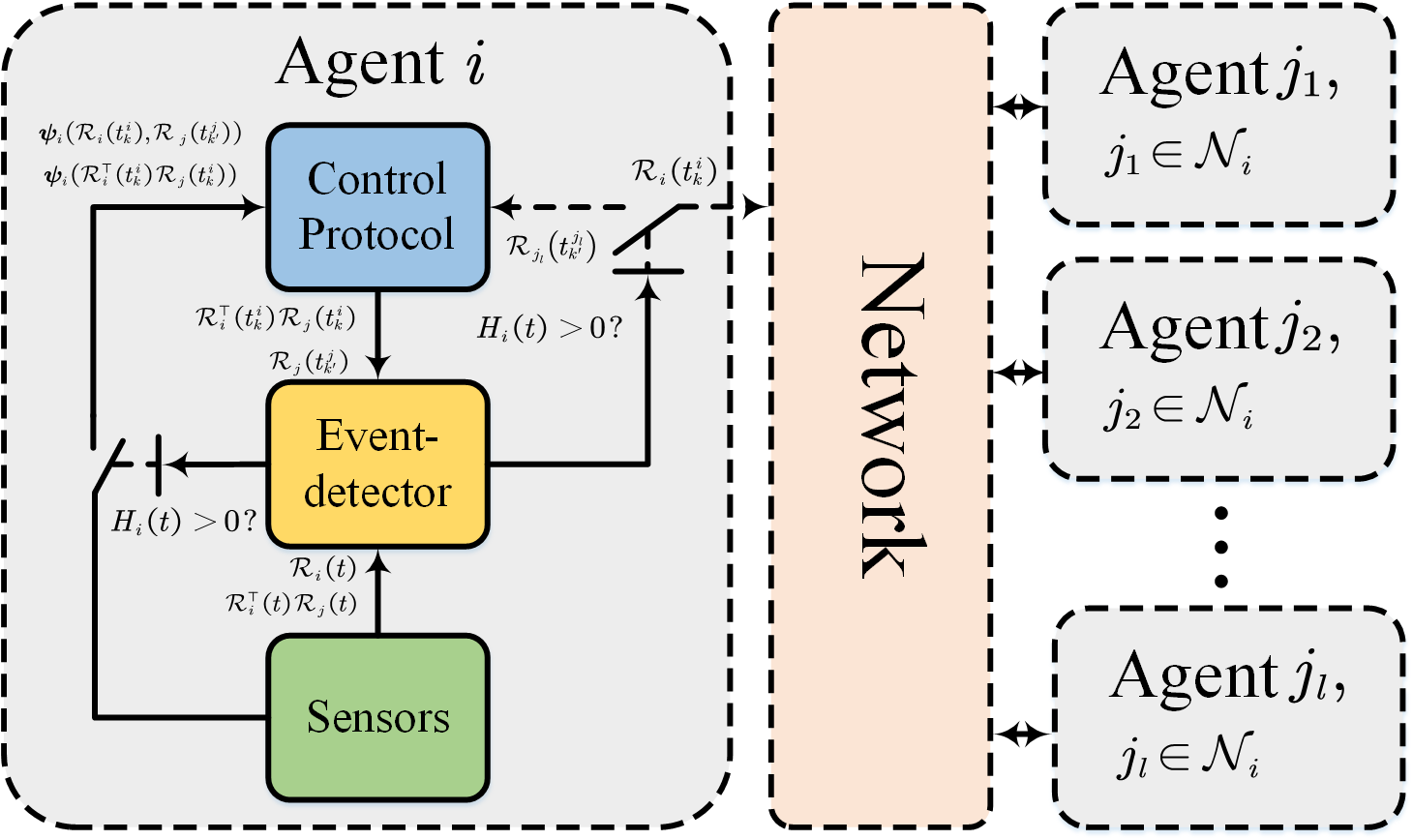}
		
		\caption{The control diagram of event-triggered attitude synchronization \cite{2020_Auto_X.Jin}.}	\label{diagram4}
	}
\end{figure}

Motivated by the limited communication resource in aerospace applications such as nano-satellite swarms, the continuous attitude synchronization protocol is not feasible. 
The event-triggered distributed control is widely investigated in the multi-agent systems \cite{2020_Chaos,2020_TAC_X.Li,2020_Auto_X.Li,2021_TAC_Y.Tang,2017_TCYB_Daniel,2023_TNNLS_Daniel,2022_Auto_Daniel} as well as rigid body systems \cite{2022_TAES_Event-triggered_attitude,2022_TNSE_X.Jin}.
An event-triggered attitude consensus is considered in the absolute attitude measurement and the relative attitude measurement cases \cite{2020_Auto_X.Jin}, respectively. 
The event-triggered attitude consensus framework is shown in Fig. \ref{diagram4}\cite{2020_Auto_X.Jin}.
\begin{table*}[htb]
	\begin{tabular}{|c|cc|cl|l|lll|c|}
		\hline
		\multirow{2}{*}{Reference} & \multicolumn{2}{c|}{Graph}                              & \multicolumn{2}{c|}{Model uncertainty}                                                                                                                                 & \multicolumn{1}{c|}{\multirow{2}{*}{Velocity-free}} & \multicolumn{3}{c|}{Communication}                                                                         & \multirow{2}{*}{\begin{tabular}[c]{@{}c@{}}Switching\\ topology\end{tabular}} \\ \cline{2-5} \cline{7-9}
		& \multicolumn{1}{c|}{Undirected} & Directed              & \multicolumn{1}{c|}{\begin{tabular}[c]{@{}c@{}}Parameter\\ linearity\end{tabular}} & \multicolumn{1}{c|}{\begin{tabular}[c]{@{}c@{}}Unmodeled\\ dynamics\end{tabular}} & \multicolumn{1}{c|}{}                               & \multicolumn{1}{c|}{Continuous} & \multicolumn{1}{c|}{Sampled-data} & \multicolumn{1}{c|}{Event-triggered} &                                                                               \\ \hline
		\cite{2009_IJC_W.Ren_EL,2010_Auto_Z.Meng_attitude}                          & \multicolumn{1}{c|}{$\checked$}           &                       & \multicolumn{1}{c|}{$\checked$}                                                              &                                                                                   &                                                     & \multicolumn{1}{c|}{$\checked$}           & \multicolumn{1}{l|}{}             &                                      &                                                                               \\ \hline
		\cite{2011_TAC_J.Mei,2012_Auto_J.Mei_EL,2013_Auto_H.Wang_EL,2014_TAC_Abd}                          & \multicolumn{1}{c|}{}           &         $\checked$              & \multicolumn{1}{c|}{$\checked$}                                                              &                                                                                   &                                                     & \multicolumn{1}{c|}{$\checked$}           & \multicolumn{1}{l|}{}             &                                      &                                                                               \\ \hline
		\cite{2015_TCNS_Dixon}                          & \multicolumn{1}{c|}{$\checked$}           & \multicolumn{1}{l|}{} & \multicolumn{1}{l|}{}                                                              &                 \multicolumn{1}{c|}{$\checked$}                                                 &                                                     & \multicolumn{1}{c|}{$\checked$}           & \multicolumn{1}{l|}{}             &                                      &                                                                               \\ \hline
		\cite{2019_Auto_Dimos}                          & \multicolumn{1}{c|}{}           & \multicolumn{1}{c|}{$\checked$} & \multicolumn{1}{l|}{}                                                              &                 \multicolumn{1}{c|}{$\checked$}                                                 &                                                     & \multicolumn{1}{c|}{$\checked$}           & \multicolumn{1}{l|}{}             &                                      &                                                                               \\ \hline
		\cite{2018_TII_W.B.Zhang,2017_TAC_Abdessameud_EL}                          & \multicolumn{1}{c|}{}           &   $\checked$                    & \multicolumn{1}{c|}{$\checked$}                                                              &                                                                                   &                                                     & \multicolumn{1}{c|}{}           & \multicolumn{1}{c|}{$\checked$}             &                                      &                                                                               \\ \hline
		\cite{2020_TCAS1,2019_TNSE_X.Jin,2022_TCST_X.Jin}                          & \multicolumn{1}{l|}{}           & \multicolumn{1}{c|}{$\checked$} & \multicolumn{1}{c|}{$\checked$}                                                              &                                                                                   &                                                     & \multicolumn{1}{l|}{}           & \multicolumn{1}{l|}{}             &      \multicolumn{1}{c|}{$\checked$}        & \multicolumn{1}{l|}{}                                                         \\ \hline
		\cite{2023_TNNLS_Y.Shi,2021_TNSE_X.Jin}                          & \multicolumn{1}{c|}{}           &         $\checked$              & \multicolumn{1}{c|}{}                                                              &                  \multicolumn{1}{c|}{$\checked$}                                                           &                                                     & \multicolumn{1}{l|}{}           & \multicolumn{1}{l|}{}             &         \multicolumn{1}{c|}{$\checked$}                                    &                                                                               \\ \hline
		\cite{2021_TAC_EL_Y.Shi}                          & \multicolumn{1}{c|}{$\checked$}           & \multicolumn{1}{l|}{} & \multicolumn{1}{c|}{$\checked$}                                                              &                                                                      &                                          \multicolumn{1}{c|}{$\checked$}                     & \multicolumn{1}{l|}{}           & \multicolumn{1}{l|}{}             &                \multicolumn{1}{c|}{$\checked$}                    & \multicolumn{1}{l|}{}                                                         \\ \hline
		\cite{2019_TAC_Abd}                          & \multicolumn{1}{c|}{}           &   $\checked$                    & \multicolumn{1}{c|}{$\checked$}                                                              &                                                                                   &                                                     & \multicolumn{1}{c|}{$\checked$}           & \multicolumn{1}{l|}{}             &                                      & \multicolumn{1}{c|}{$\checked$          }                                               \\ \hline
		\cite{2019_TAC_EL_J.Huang}                    & \multicolumn{1}{c|}{}           &     \multicolumn{1}{c|}{$\checked$}            & \multicolumn{1}{c|}{$\checked$}                                                              &                                                                                   &      \multicolumn{1}{c|}{$\checked$}                                                       & \multicolumn{1}{c|}{$\checked$}           & \multicolumn{1}{l|}{}             &                                      &                     \multicolumn{1}{c|}{$\checked$}                                               \\ \hline
		\cite{2019_TAC_G.Feng_ETC+Switching}                        & \multicolumn{1}{c|}{$\checked$}           & \multicolumn{1}{l|}{} & \multicolumn{1}{c|}{$\checked$}                                                              &                                                                                   &                                                     & \multicolumn{1}{l|}{}           & \multicolumn{1}{l|}{}             &           \multicolumn{1}{c|}{$\checked$}                        & \multicolumn{1}{c|}{$\checked$}                                                         \\ \hline
		\cite{2022_TAC_L.Liu}                        & \multicolumn{1}{c|}{}           & \multicolumn{1}{c|}{$\checked$} & \multicolumn{1}{c|}{$\checked$}                                                              &                                                                                   &                                                     & \multicolumn{1}{l|}{}           & \multicolumn{1}{l|}{}             &           \multicolumn{1}{c|}{$\checked$}                        & \multicolumn{1}{c|}{$\checked$}                                                         \\ \hline
	\end{tabular}
	{\color{black}\caption{A literature summary of coordination control of multiple rigid body systems.}	\label{review2}}
\end{table*}
In the first case, a gnomonic mapping is utilized to project the unit-quaternion on a hemi-sphere contained on $\mathbb{S}^{3}$ to the Euclidean plane almost globally. Based on the projection $\mathbf{y}_{i}=\tan\frac{\theta_{i}}{2}\mathbf{u}_{i}$, the protocol is designed as 
\begin{align}\label{event_triggered}
	\mathbf{w}_i(t)=\sum_{j=1}^N a_{i j}\left(\mathbf{y}_j\left(t_{k^{\prime}}^j\right)-\mathbf{y}_i\left(t_k^i\right)\right), \; t \in\left[t_k^i, t_{k+1}^i\right),
\end{align}
where $\mathbf{y}_i\left(t_k^i\right)$ denotes the $i$th rigid body's attitude at the triggering instant $t_{k}^{i}$.
Let the event-triggered sampling error be defined as $\mathbf{e}_{i}(t)=\mathbf{y}_{i}(t_{k}^{i})-\mathbf{y}_{i}(t)$, the triggering instant is determined by the following condition
$$
t_{k+1}^i=\min _{t \geq t_k^i}\left\{t \in \mathbb{R}:\left\|\mathbf{e}_i(t)\right\|>\alpha_i\left\|\mathbf{z}_i(t)\right\|, i=1, \ldots, N\right\},
$$
where $\mathbf{z}_i(t)=\sum_{j=1}^N a_{i j}\left(\mathbf{y}_j\left(t_{k^{\prime}}^j\right)-\mathbf{y}_i\left(t_k^i\right)\right), t \in\left[t_k^i, t_{k+1}^i\right)$ and $\alpha_i<\frac{1}{2\bar{a}|\mathcal{N}_{i}|}$.
Note that the event-triggered protocol (\ref{event_triggered}) can achieve the almost global attitude consensus due to the property that the configuration space of the projection $\mathbf{y}_{i}\in \mathcal{B}_{r}(\mathbf{0})$, where $r=\infty$. 

In the second protocol, a gradient vector for a disagreement function using the geodesic distance on $\mathbb{SO}(3)$ is utilized to achieve attitude synchronization. 
Define a geodesic distance function as follows
\begin{align}
	\phi=\sum_{i=1}^N \phi_i=\frac{1}{2} \sum_{i=1}^N \sum_{j=1}^N a_{i j} d^2\left(\mathcal{R}_i, \mathcal{R}_j\right).
\end{align}
Then, the gradient vector of the function $\phi_{i}$ at the point $\mathcal{R}_{i}$ can be calculated by $\nabla_{\mathcal{R}_i} \phi_i:=-\sum_{j=1}^N a_{i j} \log \left(\mathcal{R}_i^{\top} \mathcal{R}_j\right)$.
Based on the gradient vector, the protocol is formulated as 
\begin{align}
	\begin{aligned}
		\mathbf{w}_i^{\wedge}(t) & =-\nabla_{\mathcal{R}_i} \phi_i\left(t_k^i\right) \\
		& =\sum_{j=1}^N a_{i j} \log \left(\mathcal{R}_i^{\top} \mathcal{R}_j\right)\left(t_k^i\right), t \in\left[t_k^i, t_{k+1}^i\right) .
	\end{aligned}
\end{align} 
The event-triggered sampling error is designed on the tangent space of $\mathbb{SO}(3)$, 
\begin{align}
	E_i(t)= & \nabla_{\mathcal{R}_i} \phi_i(t)-\nabla_{\mathcal{R}_i} \phi_i\left(t_k^i\right) \nonumber\\
	= & \sum_{j=1}^N a_{i j} \log \left(\mathcal{R}_i^{\top} \mathcal{R}_j\right)\left(t_k^i\right)-\sum_{j=1}^N a_{i j} \log \left(\mathcal{R}_i^{\top} \mathcal{R}_j\right)(t),  \nonumber\\
	& t \in\left[t_k^i, t_{k+1}^i\right),
\end{align}
and {\color{black}a dynamic event-triggered condition is given as 
\begin{align}
	t_{k+1}^i= & \min _{t \geq t_k^i}\Big\{t \in \mathbb{R}: \eta_i(t)+\theta_i\left(\alpha_i\left\|\nabla_{\mathcal{R}_i}^{\vee} \phi_i(t)\right\|^2 \right.\nonumber\\
	&\left.-\left\|E_i^{\vee}(t)\right\|^2\right) \leq 0,\;
	t \in [t_k^i, t_{k+1}^i) \Big\},
\end{align}
where $\theta_i \in\left[\frac{1-\beta_i}{\lambda_i},+\infty\right)$, and $\eta_{i}$ is a dynamic variable inspired by \cite{2015_TAC_Girard_Dynamic}. The dynamics are designed as
\begin{align}{\label{dyvar}}
	\dot{\eta}_{i}(t)=-\lambda_{i}\eta_{i}(t)+\beta_{i}\Big(\alpha_{i}\Big\|\nabla^{\vee}_{\mathcal{R}_i}\phi_{{i}}(t)\Big\|^{2}-\Big\|E_{i}^{\vee}(t)\Big\|^{2}\Big),
\end{align}
where $\eta_{i}(0)>0$, $\lambda_{i}>0$, $\beta_{i}\in[0, \frac{1}{2}]$ and $\alpha_{i}\in [0,1]$ are non-negative parameters.
The proposed event-triggered scheme guarantees the positive invariance of the set $\mathcal{B}_{\frac{\pi}{2}}(\mathcal{Q})$ on $\mathbb{SO}(3)$, where $\mathcal{Q}\in \mathbb{SO}(3)$ is a rotation, and the attitude consensus can be achieved by using only relative attitude measurements. 
The further result, which considers the dynamic model of the rigid body, and the angular velocity-free attitude synchronization scheme is proposed under event-triggered mechanisms \cite{2022_Auto_X.Jin_Event}.} 

\section{Coordination control of multiple rigid body systems}
The motion of rigid bodies has total six degree of freedom, i.e., three for orientations and three for positions. It is worth noting that the orientation and the position are often coupled in practical applications of rigid bodies such as formation flying of quadrotors \cite{2018_TMech_Y.Zou,2020_TCNS_D.Zhang}.
Different from attitude synchronization which only focuses on the orientation control, the coordination control of multiple rigid body systems pays attention to the position and the orientation control coupled together. A literature summary of this section is shown in Table \ref{review2}.

\subsection{Coordination control of Euler-Lagrange systems}
The Euler-Lagrange equation is an effective method to model the dynamic of mechanical systems in terms of the energy conservation \cite{2001_mechanics}. 
In addition, it allows a unified design of rotation and translation control law coupled together \cite{2009_JGCD_applicationattitude,2009_TRO_JJE}. 
A distributed leaderless consensus problem is considered for networked Euler-Lagrange systems \cite{2009_IJC_W.Ren_EL}.
The fundamental consensus algorithm is first proposed under the undirected graph. 
Then, two consensus algorithms accounting of actuator saturation and unavailability of measurements for generalized coordinate derivatives are proposed. However, the undirected graph condition is needed.
A distributed containment control problem is considered for networked Euler-Lagrange systems under directed graphs \cite{2012_Auto_J.Mei_EL}. 
A distributed sliding mode estimator is given by
\begin{align}\label{EL_reference}
	\begin{aligned}
		& \hat{\dot{\mathbf{q}}}_{r i} \triangleq \hat{\mathbf{v}}_i-\alpha \sum_{j \in \mathcal{V}_L \cup \mathcal{V}_F} a_{i j}\left(\mathbf{q}_i-\mathbf{q}_j\right), \\
		& \hat{\dot{\mathbf{q}}}_{r i} \triangleq \hat{\mathbf{a}}_i-\alpha \sum_{j \in \mathcal{V}_L \cup \mathcal{V}_F} a_{i j}\left(\dot{\mathbf{q}}_i-\dot{\mathbf{q}}_j\right), \\
		& \hat{\mathbf{s}}_i \triangleq \dot{\mathbf{q}}_i-\hat{\dot{\mathbf{q}}}_{r i}=\dot{\mathbf{q}}_i-\hat{\mathbf{v}}_i+\alpha \sum_{j \in \mathcal{V}_L \cup \mathcal{V}_F} a_{i j}\left(\mathbf{q}_i-\mathbf{q}_j\right), \; i \in \mathcal{V}_F,
	\end{aligned}
\end{align}
where $\mathbf{q}_{i}$ and $\dot{\mathbf{q}}_{i}$ denote the vector of generalized coordinates and the vector of the derivative of generalized coordinates, respectively. $\hat{\mathbf{v}}_i$ and $\hat{\mathbf{a}}_i$ denote the estimation of the leader's velocity and acceleration, and $\alpha$ is a positive constant. 
Based on the design of the reference signal and sliding mode variable in (\ref{EL_reference}), the adaptive distributed control protocol can be given by
\begin{align}
	& \bm{\tau}_i=-K_i \hat{\mathbf{s}}_i+Y_i\left(\mathbf{q}_i, \dot{\mathbf{q}}_i, \hat{\ddot{\mathbf{q}}}_{r i}, \hat{\dot{\mathbf{q}}}_{r i}\right) \widehat{\Theta}_i, \label{EL_adaptive}\\
	& \dot{\hat{\mathbf{v}}}_i=-\beta_1 \operatorname{sgn}\left[\sum_{j \in \mathcal{V}_F} a_{i j}\left(\hat{\mathbf{v}}_i-\hat{\mathbf{v}}_j\right)+\sum_{j \in \mathcal{V}_L} a_{i j}\left(\hat{\mathbf{v}}_i-\dot{\mathbf{q}}_j\right)\right], \label{estimatior_v}\\
	& \dot{\hat{\mathbf{a}}}_i=-\beta_2 \operatorname{sgn}\left[\sum_{j \in \mathcal{V}_F} a_{i j}\left(\hat{\mathbf{a}}_i-\hat{\mathbf{a}}_j\right)+\sum_{j \in \mathcal{V}_L} a_{i j}\left(\hat{\mathbf{a}}_i-\ddot{\mathbf{q}}_j\right)\right], \label{estimatior_a} \\
	& \dot{\widehat{\Theta}}_i=-\Lambda_i Y_i^{\top}\left(\mathbf{q}_i, \dot{\mathbf{q}}_i, \hat{\dot{\mathbf{q}}}_{r i}, \hat{\dot{\mathbf{q}}}_{r i}\right) \hat{\mathbf{s}}_i, \quad i \in \mathcal{V}_F,  \label{theta}
\end{align} 
where $\widehat{\Theta}_{i}$ is the estimation of constant physical parameter $\Theta_{i}$, $Y_i\left(\mathbf{q}_i, \dot{\mathbf{q}}_i, \hat{\tilde{\mathbf{q}}}_{r i}, \hat{\dot{\mathbf{q}}}_{r i}\right)$ is the regression vector, $K_i$ and $\Lambda_i$ are symmetric positive-definite matrices, $\beta_1$ and $\beta_2$ are positive constants.
The estimators (\ref{estimatior_v}) and (\ref{estimatior_a}) are distributed finite-time observers, which provide the leader's velocity and acceleration estimation for each follower. The adaptive law (\ref{theta}) is designed based on the linear property of Euler-Lagrange models to deal with the unknown physical parameters for rigid bodies. 
Note that this framework can also solve the leader-follower and leaderless consensus problem for networked Euler-Lagrange systems.

Following this work, there are number of results on coordination control of networked Euler-Lagrange systems \cite{2013_Auto_H.Wang_EL,2012_TRO_Chopra_EL,2020_TNSE_EL}. 
A leader-follower flocking algorithm is proposed for the leader with constant velocities and time-varying velocities, respectively, to maintain a connectivity and avoid collisions \cite{2016_Auto_Meijie_Flocking}. 
The key idea of dealing with the collision avoidance and connectivity maintenance is to introduce the potential function $V_{ij}$ as follows: 1) If $\left\|\mathbf{q}_i(0)-\mathbf{q}_j(0)\right\| \geq R$, where $R$ is sensing radius of the agents. $V_{i j}$ is a differentiable nonnegative function of $\left\|\mathbf{q}_i-\mathbf{q}_j\right\|$ satisfying the conditions:
(i) $V_{i j}=V_{j i}$ achieves its unique minimum when $\left\|\mathbf{q}_i-\mathbf{q}_j\right\|$ is equal to the value $\bar{d}_{i j}$, where $\bar{d}_{i j}<R$.
(ii) $V_{i j} \rightarrow \infty$ as $\left\|\mathbf{q}_i-\mathbf{q}_j\right\| \rightarrow 0$.
(iii) $\frac{\partial V_{i j}}{\partial\left(\left\|\mathbf{q}_i-\mathbf{q}_j\right\|\right)}=0$ if $\left\|\mathbf{q}_i-\mathbf{q}_j\right\| \geq R$.
(iv) $V_{i i}=c, i=1, \ldots, n$, where $c$ is a positive constant.
(2) If $\left\|\mathbf{q}_i(0)-\mathbf{q}_j(0)\right\|<R, V_{i j}$ is defined as above except that condition (iii) is replaced with the condition that $V_{i j} \rightarrow \infty$ as $\left\|\mathbf{q}_i-\mathbf{q}_j\right\| \rightarrow R$.
Based on this potential function, the distributed algorithm is given by 
\begin{align}
	& \bm{\tau}_i=\hat{\bm{\tau}}_i+\mathbf{Y}_i\left(\mathbf{q}_i, \dot{\mathbf{q}}_i, \dot{\mathbf{v}}_i, \mathbf{v}_i\right) \hat{\theta}_i, \label{EL_controller_2} \\
	& \hat{\bm{\tau}}_i=-\sum_{j=0}^n \frac{\partial V_{i j}}{\partial \mathbf{q}_i}-\gamma \sum_{j=0}^n a_{i j}(t)\left(\dot{\mathbf{q}}_i-\dot{\mathbf{q}}_j\right), \\
	& \dot{\mathbf{v}}_i=-\sum_{j=0}^n \frac{\partial V_{i j}}{\partial \mathbf{q}_i}-\gamma \sum_{j=0}^n a_{i j}(t)\left(\dot{\mathbf{q}}_i-\dot{\mathbf{q}}_j\right), \\
	& \dot{\hat{\theta}}_i=-\Gamma_i \mathbf{Y}_i^{\top}\left(\mathbf{q}_i, \dot{\mathbf{q}}_i, \dot{\mathbf{v}}_i, \mathbf{v}_i\right) \mathbf{s}_i, \label{adaptive}
\end{align}
where $a_{ij}(t)$ is the weight associated with the proximity graph.

{\color{black}The structure of the above controllers, including (\ref{EL_adaptive}) and (\ref{EL_controller_2}) contains two parts, which are constructed based on the fundamental properties of  Euler-Lagrange equations. The first part is the coordination feedback term, which drives the states to synchronization. 
The theoretical analysis is made by designing the Lyapunov function $V=\hat{\mathbf{x}}_{i}^{\top}\mathbf{M}_{i}\hat{\mathbf{x}}_{i}$ or $V=\hat{\mathbf{q}}_{i}^{\top}\mathbf{M}_{i}^{\top}\hat{\mathbf{q}}_{i}$ and the anti-symmetric property of Euler-Lagrange dynamics. The second part is the linear regression, and the parameter adaptive law (\ref{theta}) and (\ref{adaptive}) are built to deal with the parameter uncertainty of rigid body dynamics.

However, the parameter linearity may not be satisfied for Euler-Lagrange systems with unknown dynamics. A radial basis function neural network (RBFNN) approximation technique is implemented to solve unknown dynamics \cite{2023_TNNLS_Y.Shi}. The nonlinear dynamics 
$f(\boldsymbol{x})$ is modeled as
\begin{align}
	f(\boldsymbol{x})=\boldsymbol{W}^{\top} \boldsymbol{H}(\boldsymbol{x})+\zeta(\boldsymbol{x}), \forall \boldsymbol{x} \in \Omega_{\boldsymbol{x}},
\end{align}
where $\Omega_{\boldsymbol{x}}$ is a compact set when RBFNN is used, 
$\boldsymbol{W}=\left[w_1, w_2, \ldots, w_l\right]^{\top} \in \mathbb{R}^l$ is a weight matrix, $l$ represents the node number of neural networks, and $\zeta(\boldsymbol{x})$ is the bounded approximation errors. $\boldsymbol{H}(\boldsymbol{x})=$ $\left[h_1(\boldsymbol{x}), h_2(\boldsymbol{x}), \ldots, h_l(\boldsymbol{x})\right]^{\top}$ presents a vector of radial basis function. Then, the coordination controller of Euler-Lagrange system is designed as 
\begin{align}
		{\boldsymbol{\tau}}_i & =k_i \boldsymbol{s}_i-\hat{\boldsymbol{W}}_i^{\top} \boldsymbol{H}_i \\
		\dot{\hat{\boldsymbol{W}}}_i & =-\gamma\left(\boldsymbol{H}_i \boldsymbol{s}_i^{\top}+\sigma_i \hat{\boldsymbol{W}}_i\right), \label{NN}
\end{align}	
where $k_i, \gamma, \sigma_i>0,0<\beta_i<1$. The main difference is the adaptive law (\ref{NN}), which is based on RBFNN approximation.
In addition, there are some other approaches, such as robust integral sign of the error strategy \cite{2015_TCNS_Dixon} and augmented system method \cite{2021_TNSE_X.Jin}, which are proposed to solve the dynamics uncertainties of Euler-Lagrange systems.}

\subsection{Coordination control of multiple rigid body systems on SE(3)}
A large amount of results on coordination of Euler-Lagrange systems are based on three fundamental properties of Euler-Lagrange systems in Section 2E. While it may not always be satisfied in some practical applications such as the motion evolving on non-Euclidean manifold and under the external disturbances.
The formation control problem on $\mathbb{SE}(3)$ is studied with directed and switching topologies \cite{2016_Auto_J.Thunberg}. 
The main idea is to transform the formation problem into the consensus problem of multi-agent systems. 
Let the state of each agent be described by 
\begin{align}\label{P}
	\mathbf{P}_i(t)=\left[\begin{array}{cc}
		\mathcal{R}_i(t) & \mathbf{T}_i(t) \\
		0 & 1
	\end{array}\right] \in \mathbb{S E}(3),
\end{align}
at each time $t \geq t_0$. The matrix $\mathcal{R}_i(t)$ is an element of $\mathbb{SO}(3)$, and 
the vector $\mathbf{T}_i(t)$ is an element in $\mathbb{R}^3$.
The relative transformation on $\mathbb{SE}(3)$ is given as 
\begin{align}
	\mathbf{P}_{i j}(t) & =\mathbf{P}_i^{-1}(t) \mathbf{P}_j(t) \\
	& =\left[\begin{array}{cc}
		\mathcal{R}_{ij}(t)& \mathbf{T}_{ij}(t) \\
		0 & 1
	\end{array}\right],  \nonumber
\end{align}
which contains the relative rotation $\mathcal{R}_{ij}(t)=\mathcal{R}_i^{\top}(t) \mathcal{R}_j(t) $ and the relative translation $\mathcal{R}_i^{\top}(t)\left(\mathbf{T}_j(t)-\mathbf{T}_i(t)\right)$.
Let 
\begin{align}
	\bm{\xi}_i=\left[\begin{array}{cc}
	{\bm{\omega}}_i^{\wedge} & \bm{v}_i \\
		0 & 0 \nonumber
	\end{array}\right],
\end{align}
where $\bm{\omega}_{i}$ is the angular velocity input and $\bm{v}_{i}$ is the transition velocity input.
Then, the kinematics of $\mathbf{P}_{i}$ can be formulated as 
\begin{align}
	\dot{\mathbf{P}}_{i}=\mathbf{P}_{i}\bm{\xi}_{i}.
\end{align}
The control goal of the formation is to make $\|\mathbf{P}_{i}^{\top}\mathbf{P}_{j}(t)- \mathbf{P}_{ij}^{*}\|\rightarrow 0$ as $t \rightarrow 0$, where $\mathbf{P}_{ij}^{*}$ is the desired formation pattern.
Based on the absolute and relative transformations, the control protocol can be designed as 
\begin{align}
	& \bm{\xi}_i=\sum_{j \in \mathcal{N}_i(t)} a_{i j}(t)\left(\left(\mathbf{P}_j-\mathbf{P}_{i }\right)+\left(\mathbf{P}_i^{-1}-\mathbf{P}_j^{-1}\right)\right), \label{formation1}\\
	& \bm{\xi}_i=\sum_{j \in \mathcal{N}_i(t)}a_{i j}(t)\left(\mathbf{P}_{i j}-\mathbf{P}_{i j}^{-1}\right), \label{formation2}
\end{align}
respectively.
For the absolute transformation case, when using the protocol (\ref{formation1}), the formation control can be achieved if each initial rotation is contained in $\mathcal{B}_{\frac{\pi}{2}}(\mathbf{I}_{3})$. 
For the relative transformation case, when using the protocol (\ref{formation2}), the formation control can be achieved if all initial rotations are contained in $\mathcal{B}_{q}(\mathbf{I}_{3})$, where $q<\frac{\pi}{4}$. The asymptotic convergence will be achieved by using
the protocols (\ref{formation1}) and (\ref{formation2}). 
{\color{black}In addition, recalling the knowledge in Section C, we know that the parameterized attitude representation is formulated as $f(\mathcal{R}_{i})=g(\theta_{i})\mathbf{u}_{i}(\mathcal{R}_{i})$. When the condition $g(\theta_{i})>k\theta_{i},\; k>0$ is satisfied, and the topology remains strongly connected for all $t>0$, a special result of the exponential convergence for the formation control of multiple rigid body systems will be reached where the parameter $k$ determines the exponential rate.}

The protocols (\ref{formation1}) and (\ref{formation2}) are both designed at the kinematic level. 
A robust formation control on $\mathbb{SE}(3)$ is considered with prescribed transit and steady-state performance \cite{2019_Auto_Dimos}. 
The rigid body's motion is modeled by the Euler-Lagrange equation at the dynamic level as follows:
\begin{align}
	\mathbf{M}_i \dot{\mathbf{V}}_i+\mathbf{C}_i\left(\mathbf{V}_i\right) \mathbf{V}_i+\mathbf{G}_i\left(\mathbf{P}_i\right)+\bm{w}_i\left(\mathbf{P}_i, \mathbf{V}_{i}, t\right)=\bm{\tau}_i,
\end{align}
where $\mathbf{P}_{i}\in \mathbb{SE}(3)$ is defined in (\ref{P}), $\mathbf{V}_{i}=[\bm{v}_{i}, \bm{\omega}_{i}]\in \mathbb{R}^{6}$, $\mathbf{M}_i \in \mathbb{R}^{6 \times 6}$ is the constant positive definite inertia matrix, $\mathbf{C}_i: \mathbb{R}^6 \rightarrow \mathbb{R}^{6 \times 6}$ is the Coriolis matrix, $\mathbf{G}_i: \mathbb{S E}(3) \rightarrow$ $\mathbb{R}^6$ is the body-frame gravity vector, $\bm{w}_i: \mathbb{SE}(3) \times \mathbb{R}^6 \times \mathbb{R}_{\geq 0} \rightarrow \mathbb{R}^6$ is a bounded vector representing model uncertainties and external disturbances, and $\bm{\tau}_i \in \mathbb{R}^6$ is the control input representing the $6 \mathrm{D}$ body-frame generalized force acting on rigid body $i$. 
Supposed that the desired formation pattern is specified by $d_{k, \; \text{des}}\in\mathbb{R}^{3}$, $\mathcal{R}_{k, \text { des }}\in \mathbb{SO}(3)$, $\forall k \in \mathcal{K}$, the control objective is to design a distributed control input $\bm{\tau}_{i}\in \mathbb{R}^{6}$ such that the following requirements are satisfied $\forall k_{1}, k_{2}\in \mathcal{K}$:  
1) $\lim _{t \rightarrow \infty}\left\|\mathbf{T}_{k_2}(t)-\mathbf{T}_{k_1}(t)\right\|=d_{k, \text{des}} ;$ 
2) $\lim _{t \rightarrow \infty}\left[\mathcal{R}_{k_2}(t)\right]^{\top} \mathcal{R}_{k_1}(t)=\mathcal{R}_{k, \text { des }} ;$ 
3) $d_{k, \text { col }}<\| \mathbf{T}_{k_2}(t)-$ $\mathbf{T}_{k_1}(t) \|<d_{k, \text { con }}, \forall t \in \mathbb{R}$, where $d_{k, \text { col }}$ is the safe distance and $d_{k, \text { con }}$ is the sensing distance between rigid bodies \cite{2019_Auto_Dimos}.
In addition to the above requirements, there exist constraints of geometric topology of attitude configuration space.   
To guarantee all control objectives, by transforming the requirements into state constraints, the prescribed performance control is utilized to design the control input of rigid body systems \cite{2019_Auto_Dimos}.  
Note that the desired formation defined by orientation and distance $d_{k, \; \text{des}}\in\mathbb{R}^{3}$, $\mathcal{R}_{k, \text {des}}\in \mathbb{SO}(3)$, $\forall k \in \mathcal{K}$ is not guaranteed to be a rigidity formation, which means that the formation cannot be uniquely determined. 
To solve this problem, the bearing rigidity theory can be utilized to derive the condition that the formation can be uniquely determined up to a transition and inter-neighbor bearings with a scaling factor \cite{2020_TAC_Ahn}.
A necessary and sufficient condition for the bearing rigidity is extended to the manifold such as $\mathbb{SE}(3)$, and the heterogeneous agent dynamics on different manifolds such as $\mathbb{S}^{1}$ and $\mathbb{SO}(3)$ are also considered \cite{2021_TCNS_D.Zelazo}. 
Based on the bearing rigidity condition, numerous studies have been conducted on the bearing-based formation control for multi-agent systems \cite{2016_TAC_S.Zhao,2023_TAC_M.Cao,2019_TAC_S.Zhao,2015_TCNS_S.Zhao}. 
\subsection{Networked coordination control of multiple rigid body systems}
The above result assumes that the communication environment is ideal and reliable. 
The leader-follower consensus problem of networked Euler-Lagrange systems is studied under constrained communication  \cite{2017_TAC_Abdessameud_EL}.
The constrained communication implies that the communication between agents can be intermittent, which is subject to irregular communication time delays and packet dropouts.
For each follower $i \in \mathcal{V}_{F}$, the control input is given by 
\begin{align}
	\begin{aligned}
		\boldsymbol{\tau}_i & =\mathbf{Y}_i \hat{\Theta}_i-k_i^s\left(\dot{\mathbf{q}}_i-\dot{\mathbf{q}}_{r_i}\right), \\
		\dot{\hat{\Theta}}_i & =-\Pi_i \mathbf{Y}_i^{\top}\left(\dot{\mathbf{q}}_i-\dot{\mathbf{q}}_{r_i}\right),
	\end{aligned}
\end{align}
where $k_{i}^{s}$ is a positive constant and $\Pi_i$ is a symmetric positive-definite matrix. 
The reference velocity signal $\dot{\mathbf{q}}_{r_i}\in \mathbb{R}^{n}$ is designed as follows
\begin{align}\label{qri}
	\left[\begin{array}{c}
		\dot{\mathbf{q}}_{r_i} \\
		\dot{\hat{\mathbf{v}}}_i
	\end{array}\right]=\mathbf{S}\left[\begin{array}{l}
		\mathbf{q}_i \\
		\hat{\mathbf{v}}_i
	\end{array}\right]+\bm{\eta}_i \quad i \in \mathcal{V}_\mathcal{F},
\end{align}
where $\mathbf{S}\in\mathbb{R}^{2n\times 2n}$ is the system matrix of the leader's dynamic model, and $\bm{\eta}_{i}\in \mathbb{R}^{2N}$ is an input which is designed as 
\begin{align}\label{eta}
	\bm{\eta}_i=-k_{p_i}\left(\mathbf{x}_i-\bm{\varphi}_i\right)
\end{align}
with	
\begin{align}
	\dot{\bm{\varphi}}_i & =\mathbf{S} \bm{\varphi}_i-k_{d_i}\left(\bm{\varphi}_i-\bm{\psi}_i\right) \label{phi1}\\
	\dot{\bm{\psi}}_i & =\mathbf{S} \bm{\psi}_i-k_{{\psi}_i}\left(\bm{\psi}_i-\frac{1}{\kappa_i} \sum_{j=1}^{n+1} a_{i j} \mathbf{x}_{i j}^*\right), \label{phi2}
\end{align}
where $\mathbf{x}_{i}=[\mathbf{q}_{i}^{\top}, \hat{\mathbf{v}}_{i}^{\top}]^{\top}$, $k_{p_i}, k_{d_i}, k_{\psi_i}>0$ are scalar gains, $\kappa_i:=\sum_{j=1}^{n+1} a_{i j}$, and 
\begin{align}\label{prediction}
	\mathbf{x}_{i j}^*(t):=e^{\mathbf{S}\left(t-t_{k_{i j}}^{\mathrm{m}}(t)\right)} \mathbf{x}_j\left(t_{k_{i j}^{\mathrm{m}}(t)}\right),
\end{align}
where $\mathbf{x}_j\left(t_{k_{i j}^{\mathrm{m}}(t)}\right)$ is the most recent information of agent $j\in \mathcal{N}_{i}$ transmitted to agent $i$. Namely, $\mathbf{x}_{i j}^*(t)$ is the approximation or prediction of the $j$th agent for $i$th agent.
One of benefits of the above control framework is that it can guarantee the continuity of the control input $\bm{\tau}_{i}$. The stability of the above dynamic system (\ref{qri}), (\ref{phi1}), and (\ref{phi2}) can be shown by using the small gain theorem. 

{\color{black}Following the main idea from the above work \cite{2017_TAC_Abdessameud_EL}, the consensus problem of networked Euler-Lagrange systems is further studied under jointly connected topologies \cite{2019_TAC_Abd}. The main problem is to design the reference velocity input under switching typologies. A high-order dynamic system is constructed in the following form,
\begin{align}
	& \dot{\mathbf{x}}_i=\left(\mathbf{A}_i \otimes \mathbf{I}_m\right) \mathbf{x}_i+\left(\mathbf{B}_i \otimes \mathbf{I}_m\right) \mathbf{u}_i \\
	& \mathbf{y}_i=\left(\mathbf{C}_i \otimes \mathbf{I}_m\right) \mathbf{x}_i,
\end{align}
where $\mathbf{x}_i \in \mathbb{R}^{k m}$, for some $k>0, \mathbf{y}_i \in \mathbb{R}^m$, $\mathbf{A}_i \in \mathbb{R}^{k \times k}$, $\mathbf{B}_i \in \mathbb{R}^k$, and $\mathbf{C}_i \in \mathbb{R}^{1 \times k}$, as well as the input $\mathbf{u}_i \in \mathbb{R}^m$.
The high-order system is a high-order filter system where the input $\mathbf{u}_{i}$ involves the intermittent information transmitted from the switching neighboring agents.
Through the filter system, the reference velocity is given by $\mathbf{v}_i\triangleq-\gamma_i\left(\mathbf{q}_i-\mathbf{y}_i\right)$, which ensures a continuously differentiable torque input.

Motivated by the limited communication resource in practical application of coordination control of autonomous systems, the research on event-triggered coordination control of multiple rigid body systems has drawn growing attention. The challenging primary lies in introducing the event-triggered sampling in inherent nonlinear rigid body dynamics. The rigid body dynamics are naturally continuous. The event-triggered control turns the closed-loop dynamics into hybrid dynamics, which brings the technical difficulty in revealing the convergence performance. An event-triggered formation control protocol is proposed based on a similar framework in Section 4A, while the event-triggered control is designed based on Barbalat’s lemma and small-gain theorem \cite{2022_TCST_X.Jin}. The sampling-induced errors are regarded as disturbances by using ISS (Input-to-state stability) in the convergence analysis. 
Another interesting problem is the event-triggered coordination control of multiple rigid body systems under jointly connected topologies.
Since the topologies may switch during the inter-event interval, the inconsistency between the protocol and the current topologies should be tackled \cite{2022_TAC_L.Liu}. 
In addition, the switching topologies also bring additional difficulties in excluding Zeno bebavior due to the triggering condition being related to the switching instants \cite{2019_TAC_G.Feng_ETC+Switching}. }

\begin{figure}[h]	
	\centering
	{
		\includegraphics[scale=0.3]{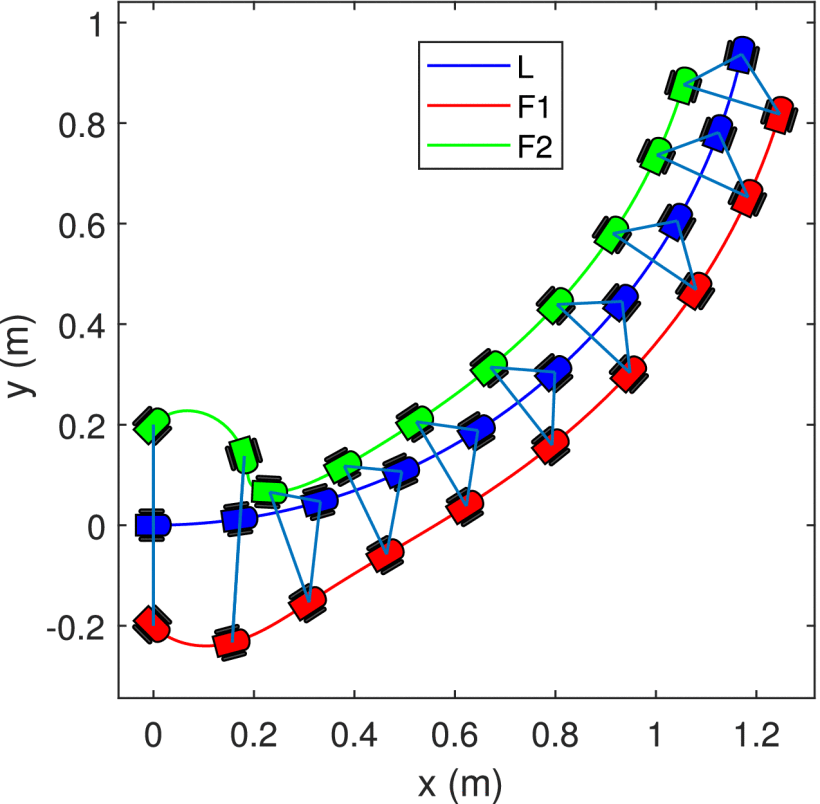}
		
		\caption{Formation trajectories of the three agents  \cite{2020_TMech_Z.Sun}.}
	\label{mobile1}}
\end{figure}
\begin{figure}[htb]	
	\centering
	{
		\includegraphics[scale=0.4]{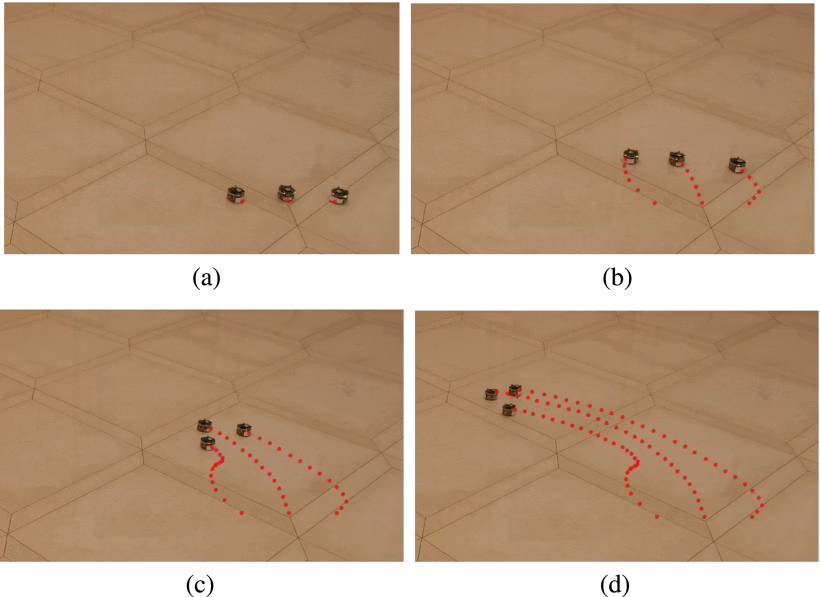}
		\caption{Experiment snap of formation control for three mobile robots \cite{2020_TMech_Z.Sun}. (a). t= s. (b) t=5s. (c) t=13s. (d) t=26s.}	\label{mobile2}
	}
\end{figure}	
\begin{figure}[htb]	
\centering
{
	\includegraphics[scale=0.3]{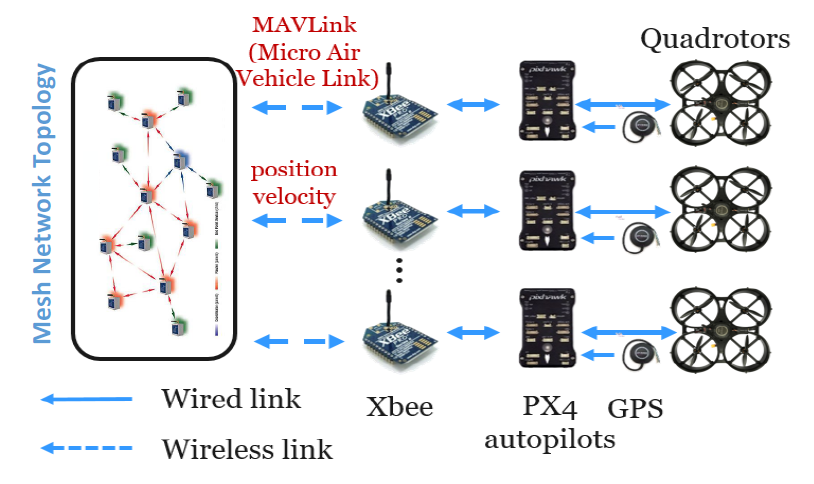}
	
	\caption{Experiment platform of formation control for three quadrotors\cite{2022_TCST_X.Jin}.}	
	\label{wireless}
}
\end{figure}
\subsection{Experiment results on coordination control of multiple rigid body systems}
\begin{figure*}[htb]	
	\centering
	\subfigure[]	
	{
		\includegraphics[scale=0.5]{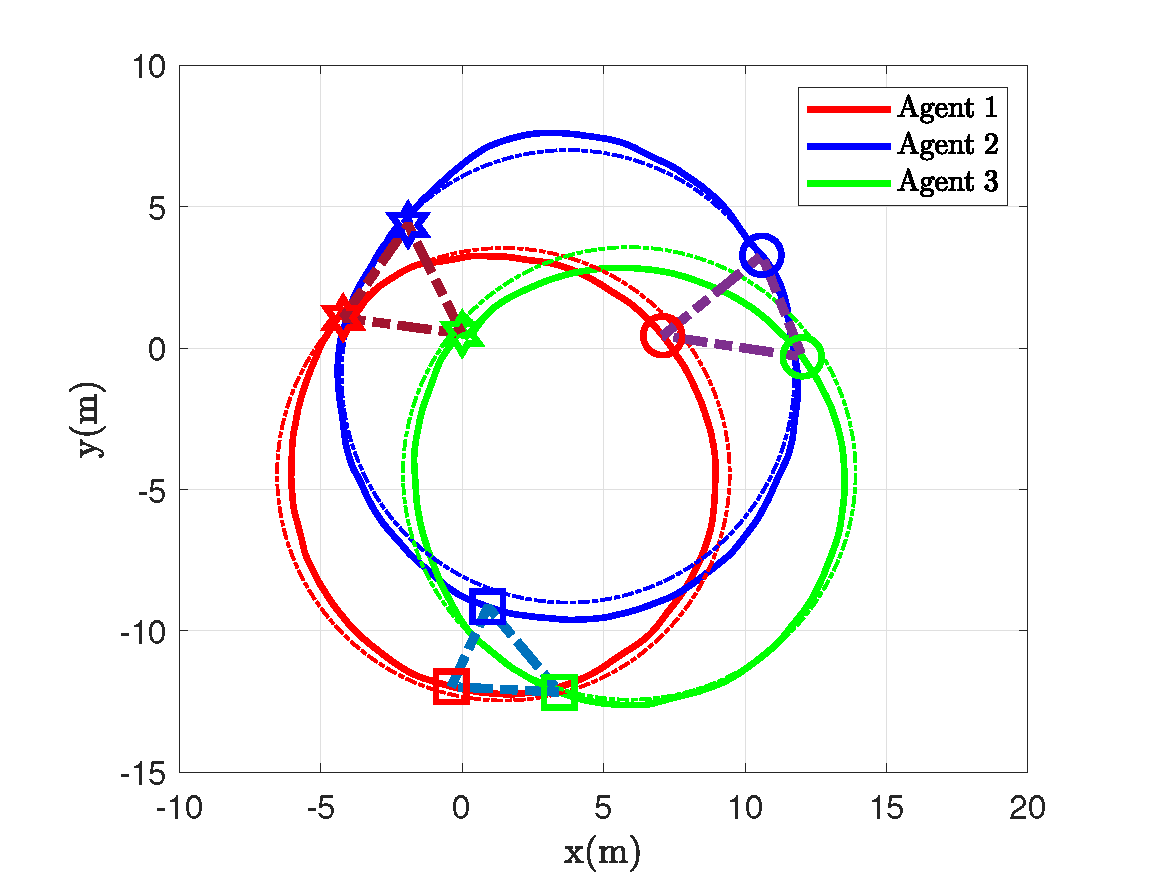}
		\label{s1}
	}	
	\subfigure[]
	{
		\includegraphics[scale=0.5]{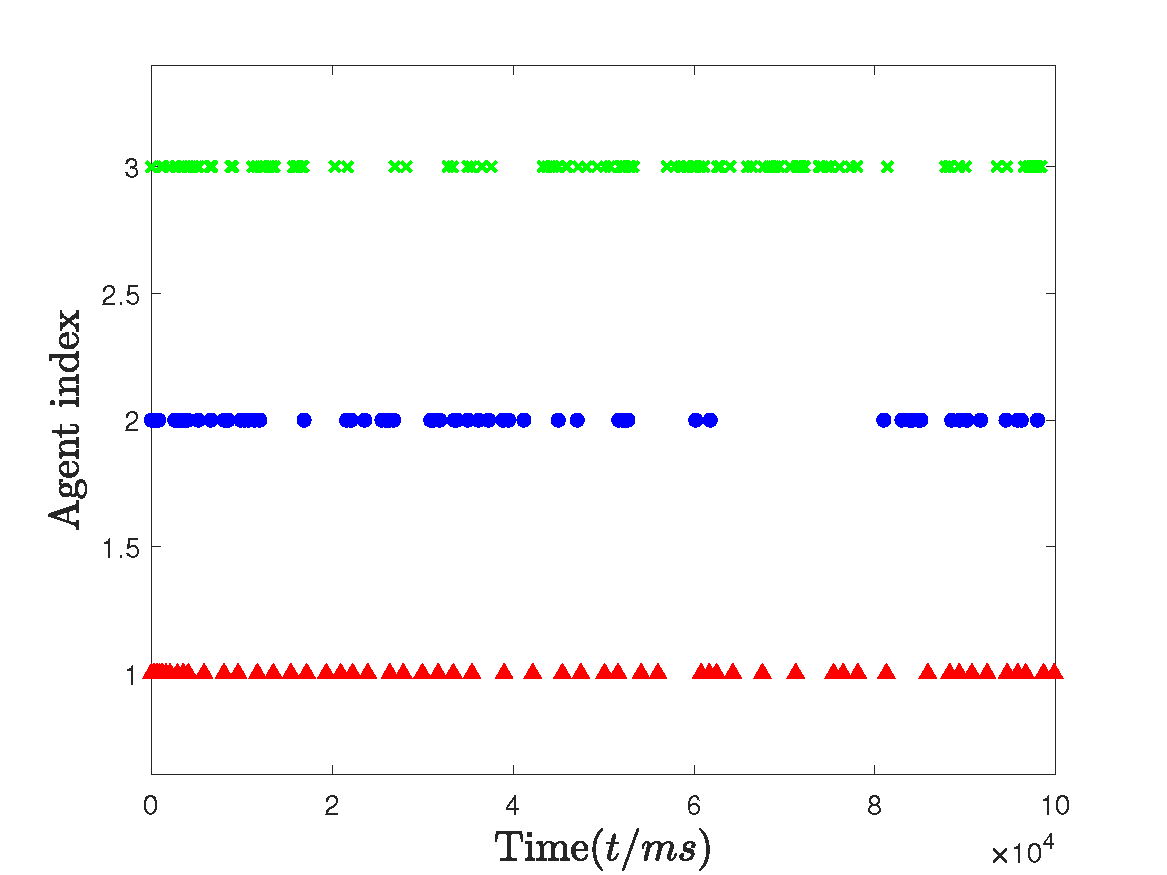}
		\label{s2}
	}	
	\caption{The experiment results of formation control of multiple quadrotors. (a). Formation trajectories of the three quadrotors. (b). Trigger instants of the three quadrotors\cite{2022_TCST_X.Jin}.}		
\end{figure*}
The coordination control of multiple rigid body systems has wide applications in robotics, transportation, and aerospace.
Therefore, researches on practical experiments of coordination control of multiple rigid bodies have been extensively conducted in the literature, including formation control of mobile robots \cite{2015_TCST_Mobile}, unmanned aerial vehicles \cite{2015_TCST_XiwangDong_Formationcontrolfouav}, and unmanned surface vehicles \cite{2021_TII_Z.Peng}. Here, we only introduce some representative results.

A trajectory tracking and mobile formation coordination on $\mathbb{SE}(3)$ is considered for a group of non-holonomic vehicles \cite{2020_TMech_Z.Sun}.
A desired mobile formation control is studied with motion constraints,  including weak rigid body motion and strict rigid body motion. The mobile formation under a weak rigid body motion preserves the relative position with the world frame while the mobile formation under a strict rigid body motion preserves the relative position as well as relative attitudes. 
In Fig. \ref{mobile1}, three robots maintain a mobile formation with strict rigid body motion.
To show the applicability of the formation control algorithm, a real experiment using three non-holonomic robots named wheeled E-puck robots is demonstrated \cite{2020_TMech_Z.Sun}.
The communication link among three robots is a directed graph, and the frequency of the communication is 0.1s through blacktooth.
The real-time trajectories of three robots in the experiment are shown in Fig. \ref{mobile2}. 
It shows that three robots form a strict rigid body motion similar to the numerical result.

{\color{black}Unmanned aerial vehicles' outdoor formation control is studied in multi-agent frameworks, where a time-varying formation is demonstrated under a switching topology in \cite{2017_X.Dong_Time}.  
In some applications of formation control of unmanned aerial vehicles, the communication resource and energy are limited.  
An event-triggered formation control is considered for Euler-Lagrange systems \cite{2022_TCST_X.Jin}. 
The quadrotor dynamic model is formulated based on the Euler-Lagrange equation.
		The inner-outer loop control strategy is proposed to achieve the formation control of quadrotors, where the inner loop control is utilized to stabilize the attitude, and the outer loop control guarantees the position and velocity tracking.
		Thus, to verify the effectiveness of the formation control algorithm, an outdoor formation experiment with three quadrotors is conducted. 
		The flight control experimental system is illustrated in Fig. \ref{wireless}.}
	
		The data transmission between quadrotors is built by the DIGI Xbee communication module. 
		The position and velocity information of each quadrotor is obtained by the GPS module. 
		The proposed control algorithm runs on Pixhawk open-source flight controller, which also integrates accelerometers and gyroscope sensors. 
		The experiment result is shown in Fig. \ref{s1}.
		In the experiment, the communication among quadrotors is governed by an event-triggered broadcasting strategy, which means that each quadrotor only broadcasts its position and velocity information to its neighbor at the triggering instants.
		The triggering instants are marked by the circle, triangle, and cross in Fig. \ref{s2}. 
		It can be observed that the communication frequency is much reduced compared with the periodic communication strategy. 
		
\begin{figure}[htb]	
			\centering
			{
				\includegraphics[scale=0.4]{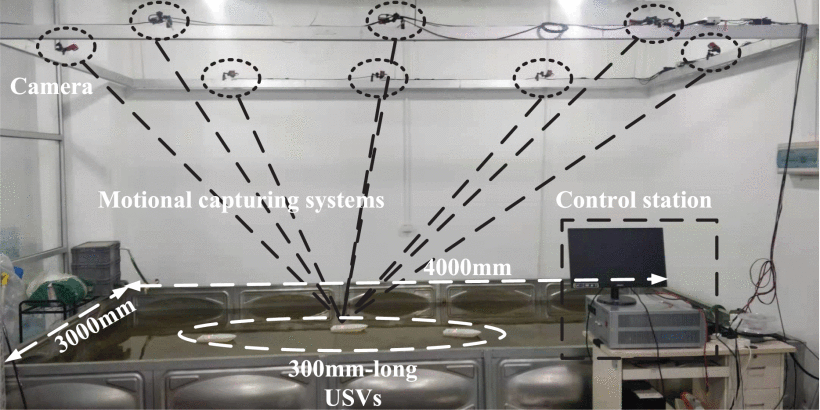}
				
				\caption{Multiple USVs control platform \cite{2023_TIE_J.W}.}	\label{USV1}
			}
		\end{figure}

\begin{figure}[htb]	
			\centering
			{
				\includegraphics[scale=0.4]{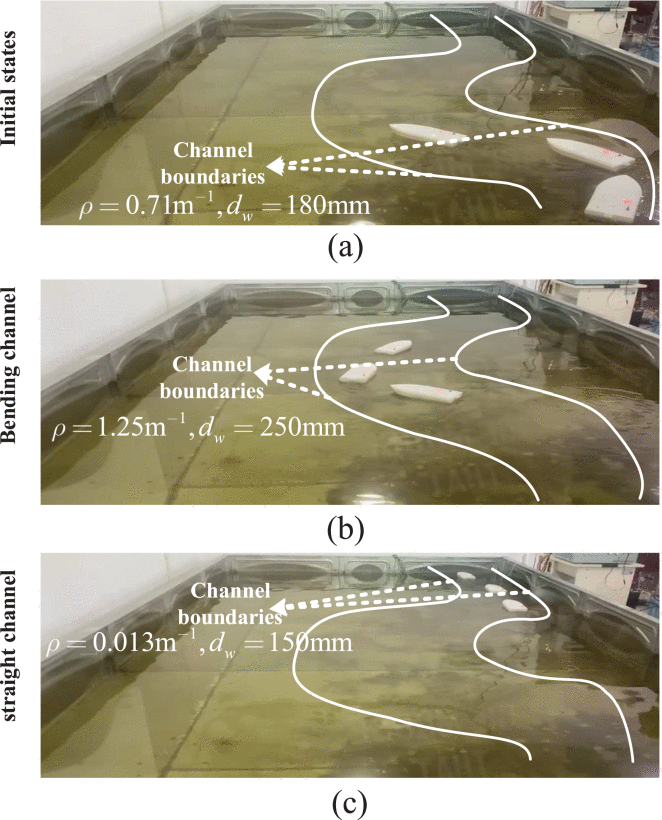}
				
				\caption{Experiment snap of formation control for three USVs. (a). Initial states. (b). Bending channel. (c). Straight channel. \cite{2023_TIE_J.W}.}	\label{USV2}
			}
		\end{figure}

		The last category of practical applications of coordination control is the formation control of unmanned vehicles, including surface vehicles and underwater vehicles. 
	An adaptive formation control of USVs is studied for navigating through narrow channels with unknown curvatures	\cite{2023_TIE_J.W}. 
		A formation tracking control protocol combing with an unknown water channel curvatures observer is proposed for steering USVs to navigate through irregular narrow channels smoothly. 
		In addition, an experiment result is also conducted based on a multiple USVs control platform which consists of a motional capturing system, a control station, 2.4-GHz wireless modules, and three USVs. 
		The platform is shown in Fig. \ref{USV1}.
		Three snapshots of the real experiment are shown in Fig. \ref{USV2}, which demonstrates a USV formation traveling along a straight channel segment and an irregular channel segment.
		It is shown that a flexible formation tracking performance can be achieved with geographical constraints of the winding channel. 

	\section{Conclusion}
    In this paper, some important topics on synchronization of multiple rigid body systems are
	surveyed from two aspects. The attitude synchronization problem of multiple rigid body systems is discussed in Section 3, where the main results are divided into local attitude synchronization, global attitude synchronization, and almost global attitude synchronization. More generally, the results of consensus on non-Euclidean spaces such as $\mathbb{S}^{n}$ and $\mathbb{SO}(n)$ are also discussed. 
	In Section 4, the coordination control of multiple rigid body systems which considers the rotational and translational motion in a unified framework is reviewed. 
	The early works on coordination control of Euler-Lagrange systems are firstly introduced based on the fundamental properties of Euler-Lagrange dynamics. 
	Then, an important topic, which considers the formation control of multiple rigid body systems on $\mathbb{SE}(3)$, and the most recent results on networked coordination control of multiple rigid body systems are discussed. 
	To verify the applicability of proposed coordination algorithms, several experimental results on the formation control of mobile robots, unmanned aerial vehicles, and unmanned surface vehicles are shown, respectively.  
	
	In the past few decades, there are much progress in synchronization of multiple rigid body systems. However, there are still some important and challenging issues that should be studied further. Some examples are listed as follows:
	\begin{itemize}
		\item   Attitude synchronization with state constraints: There are fewer existing results on attitude synchronization in presence of constraints,  which implies that there is unfeasible rotation regions. This problem is highly motivated in the scenarios that the rigid body should avoid lie in certain attitude configurations, such as undesired equilibrium points in the closed-loop dynamics or limited sensing regions in the aerospace application.  
		Due to the non-Euclidean property of $\mathbb{SO}(3)$, the synchronization protocol with state constraints on $\mathbb{SO}(3)$ is very difficult to be designed.
		
		\item  Prescribed-time attitude synchronization with absolute and relative attitude measurements: In aerospace applications, the convergence time is an important control index. It is desirable that some space tasks, such as rendezvous and docking of spacecraft, can be completed in a predefined time. However, the existing works of prescribed-time control only focus on linear models. The prescribed-time attitude synchronization, especially only relying on the relative attitude information is still an open problem.
		
		\item  Coordination control of multiple rigid body systems with communication constraints: 
		The communication network is fundamental for coordination control of multiple rigid body systems. However, in the application of underwater vehicles and spacecraft, the communication among agents is unreliable, and communication capacity is also limited which  motivates the study of the coordination control of multiple rigid body systems under communication constraints 
		
		\item A unified framework of sensing, decision-making, and control of multiple rigid body systems: Most of the existing results consider the control, estimation, and decision-making of multiple rigid body systems separately. However, in practical applications, these processes are generally coupled. The decision-making loop is dependent on the information integrated from environment sensing. The control loop is dependent on the output of the decision-making loop, and is also affected by the estimation error from the sensing loop. Hence, it is necessary to build a unified framework, which can analyze the estimation, decision-making, and control together.   
		
		\item Experiment researches on the coordination of multiple rigid body systems: Up to our knowledge, the experiment results on coordination control of multiple rigid body systems are quite limited. Most of the researches give the numerical simulation to verify the effectiveness of the proposed algorithm due to the difficulty of real experiments in the extreme environment such as deep spaces. Thus, novel verification approaches such as the semi-physical simulation are one of future directions. 
	\end{itemize}

\begin{acknowledgments}
This work was supported by the National Natural Science Foundation
of China under Grant No. 62233005, the Young Elite Scientist Sponsorship Program by Cast under Grant No. YESS20220198, the Shanghai Sailing Program under Grant No. 23YF1409600, the Hong Kong Special Administrative Region, China RGC General Research Fund under Grant CityU 11203521 and Grant CityU 11213023, the Sino-German Center for Research Promotion under Grant
M-0066, and the Program of Introducing Talents of Discipline to Universities (the 111 Project) under Grant B17017.

We wish to acknowledge Prof. J\"{u}rgen Kurths' many and groundbreaking contributions in the field of nonlinear dynamics, synchronization, and networks, and celebrate Prof.  J\"{u}rgen Kurths' $70$th birthday.

\end{acknowledgments}
%
%
%
%
%
%

\bibliography{aipsamp}

\end{document}